\documentclass[twocolumn]{aastex6}

\usepackage{aas_macros}
\usepackage{amsfonts}
\usepackage{amsmath}
\usepackage{graphicx}

\shorttitle{Dynamical Masses of Young Stars II}
\shortauthors{Rizzuto et al.}

\begin{document}
\title{Dynamical Masses of Young Stars II: Young Taurus Binaries Hubble~4, FF~Tau, and HP~Tau/G3}

\author{Aaron C. Rizzuto\altaffilmark{1,}\altaffilmark{2},
Trent J. Dupuy\altaffilmark{3},
Michael J. Ireland\altaffilmark{4},
Adam L. Kraus\altaffilmark{1}
%Sarah Maddison\altaffilmark{5}(Pending confirmation)
}
\altaffiltext{1}{Department of Astronomy, The University of Texas at Austin, Austin, TX 78712, USA}
\altaffiltext{2}{51 Pegasi b Fellow}
\altaffiltext{3}{Gemini Observatory, 670 N. Aohoku Place, Hilo, HI 96720, USA}
\altaffiltext{4}{Research School of Astronomy \& Astrophysics, Australian National University, Canberra, ACT 2611, Australia}
%%\altaffiltext{5}{Centre for Astrophysics and Supercomputing, Swinburne University of Technology, Hawthorn, Victoria 3122, Australia}

%\author{Aaron C. Rizzuto}
%\affiliation{Department of Astronomy, The University of Texas at Austin, Austin, TX 78712, USA}
%\author{Trent J. Dupuy}
%\affiliation{Gemini Observatory, 670 N. Aohoku Place, Hilo, HI 96720, USA}
%\author{Adam L. Kraus}
%\affiliation{Department of Astronomy, The University of Texas at Austin, Austin, TX 78712, USA}
%\author{Michael J. Ireland}
%\affiliation{Research School of Astronomy \& Astrophysics, Australian National University, Canberra, ACT 2611, Australia}

\begin{abstract}
One of the most effective ways to test stellar evolutionary models is to measure dynamical masses for  binary systems at a range of temperatures. In this paper, we present orbits of three young K+M binary systems in Taurus (Hubble~4, FF~Tau, and HP~Tau/G3) with VLBI parallaxes. We obtained precision astrometry with Keck-II/NIRC2, optical photometry with  HST/WFC3, and low-resolution optical spectra with WIFeS on the ANU 2.3\,m telescope. We fit orbital solutions and dynamical masses with uncertainties of 1-5\% for the three binary systems. The spectrum, photometry, and mass for Hubble~4 are inconsistent with a binary system, suggesting that it may be a triple system where the primary component consists of two stars.  For HP~Tau/G3 and FF~Tau, model masses derived from SED determined component temperatures and luminosities agree with the dynamical masses, with a small offset towards larger model masses. We find model ages for the primary components of these systems of $\sim$3\,Myr, but find that the secondaries appear younger by a factor of two. These estimates also disagree with the age of the physically associated G-type star HP~Tau/G2, which is older ($\sim$5\,Myr) according to the same models.  This discrepancy is equivalent to a luminosity under-prediction of 0.1-0.2\,dex, or a temperature over-prediction of 100-300\,K, for K/M-type stars at a given model age. We interpret this as further evidence for a systematic error in pre-main sequence evolutionary tracks for convective stars.  Our results reinforce that the ages of young populations determined from the locus of M-type members on the HR-diagram may require upward revision.

\end{abstract}
\keywords{stars: young, stars: fundamental parameters, techniques: interferometric, instrumentation: adaptive optics, instrumentation: high angular resolution}

\maketitle

\section{Introduction} 
\label{secIntro}

\begin{table*}
\centering
\caption{Properties of the three Taurus binary systems. }
\label{tprops}
\begin{tabular}{ccccccccc}
\hline
\hline
2MASS & Name & R.A. & Decl. & SpT & r' & K & $\pi$ \\
	     &             &  (J2000) & (J2000)& & (mag) & (mag) &(mas)\\
\hline
J04352089+2254242 & FF~Tau        & 04 35 20.90 & +22 54 24.3 &  K8  & 13.1  & 8.59   & (6.20$\pm$0.03)\\
J04184703+2820073 & Hubble~4    & 04 18 47.04  & +28 20 07.3 & K8.5   & 12.0  & 7.29   & 7.686$\pm$0.032\\
J04355349+2254089 & HP~Tau/G3 & 04 35 53.50    & +22 54 09.0   & M0.6   & ...      & 8.80   & (6.20$\pm$0.03)\\
J04355415+2254134 & HP~Tau/G2 & 04 35 54.15    & +22 54 13.6 & G2   & 10.6  & 7.23   & 6.20$\pm$0.03\\
\hline
\end{tabular}
\tablecomments{Spectral types are taken from \citet{herczeg14}, r' and K band magnitudes are taken from APASS \citep{apass} and 2MASS \citep{2mass} with typical uncertainties of 0.1 and 0.02\,mags respectively. System parallaxes are from the VLBI observations of \citet{Torres07,Torres09} for HP~Tau/G3 and FF~Tau, and from the latest observations of \citet{galli18}. The parallaxes for FF Tau and HP Tau/G3 are given in parentheses because we are adopting the parallaxes measurement from the associated and bound star HP~Tau/G2.}
\end{table*}

Age dated stellar populations establish the timeline for the study of many different astrophysical processes, including disk evolution and dissipation (e.g., \citealt{rieke05,carpenter06,carpenter09,chen11,rizzuto12,luhman12}), exoplanet formation and migration (e.g., \citealt{krilandlkca15,zeit1,zeit3,trevor_k233,donativ830}) and stellar gyrochronology (e.g., \citealt{mamajek08,douglas16}). In the absence of associated evolved high-mass stars to map the main sequence turn-off, the descent of young low-mass stars onto the main sequence is the most sensitive tool available for dating a young ($\la$20\,Myr) association. For typical initial conditions, a solar-mass young star contracts to within 20\% of its Zero-Age Main Sequence (ZAMS) over $\sim$30\,Myr from an initial radius of more than twice its ZAMS radius - a difference in radius that is relatively easy to detect as an excess luminosity above the main sequence on a traditional HR-diagram. This sensitivity does not necessarily translate to accuracy.  Indeed there are theoretical suggestions \citep[e.g.,][]{baraffe12} that there are a range of values for the internal stellar entropy at the conclusion of accretion, as well as a range in initial rotation rates \citep[e.g.,][]{mamajek08}. However, observational evidence suggests that the majority of binary systems appear highly coeval \citep[e.g.,][]{Kraus_hill09}, with only a small minority showing measurable age differences.  Furthermore, in young groups containing early-type stars that have begun turning off the main sequence, HR-diagram position does not accurately translate to age for convective stars with the current evolutionary models \citep[e.g.,][]{soderblom10,kraususcoctio5,jeffries16,feiden16_usco}.

Multiple-star systems have been a key testing ground for pre-main sequence models \citep[e.g][]{simon13,schaefer14,montet15,schaefer16,nielsen16}. For the most part, stars in binary systems appear to be the same age, although a significant minority ($\sim$1/3) of very young ($\la$3\,Myr) systems show significant age discrepancy between their components \citep[e.g.,][]{Kraus_hill09}. Disentangling dispersion in initial conditions from uncertainties in evolutionary models and real age dispersion within a cluster requires additional data beyond temperature and luminosity. The most readily observable quantity is the dynamical mass,  which can be observed through the orbits of binary stars \citep[e.g.,][]{boden12,dupuy17} or resolved line emission measurements of gaseous circumstellar disks \citep[e.g.,][]{simon2000,czekala16,sheehan19}.

With the release of  Gaia  parallax measurements \citep{gaiadr2_arxiv} for the majority of young ($<$20\,Myr) G/K/M-type stars in the wider solar neighborhood (200\,pc), we have begun obtaining high-angular resolution monitoring of a large sample of young binary systems in star-forming regions and young associations. This campaign will build a calibration sample for the next generation of models with dynamical mass measurements at the level of the expected Gaia parallax uncertainties \citep{dmys1}. These measurements will also allow interpretation of the Gaia photocenter motion data for these young binaries, which will be contaminated by significant stellar variability (e.g. \citealt{zeit5}).

In this study, we present the orbits of three close binary systems in the Taurus-Auriga star forming region discovered during the survey of \citet{bdd2}, which have parallaxes measured with very long baseline interferometry (VLBI)  or are associated with objects that have VLBI parallaxes. In Section \ref{sample} we describe the three Taurus binary systems, in Sections \ref{keckobs} and \ref{sectorbs} we describe the NIRC2 aperture masking observations and orbit fits to the resulting astrometry, and in Section \ref{secthubble} and \ref{wifes_section} we describe the analysis of the HST photometry and WiFeS spectra, in Section \ref{sectSED} we fit two component SEDs to the binary systems to determine luminosities and temperatures, and then fit evolutionary models to the data. In Section \ref{sectmodmod} we discuss the performance of the models, and in Section \ref{sectimplications} we discuss the implications of the results on the age of the Taurus population.

\begin{table*}
\centering
\caption{Table of Keck/NIRC2 Non-Redundant Masking Observations.}
\label{nirc2tab}
\begin{tabular}{cccccccc}
\hline
\hline
Epoch & MJD & Filter & Sep  & P.A.  & Contrast \\
&&& (mas) & (deg) & (mag) \\
\hline
& & & {\bf FF~Tau} & & \\
\hline
2007-11-23 & 54427.579& K' & 36.1$\pm$0.4 & 356.4$\pm$0.5 & 1.03$\pm$0.02 \\
2008-12-21 & 54821.520 & K' & 22.6$\pm$2.2 & 342.0$\pm$2.0 & 1.96$\pm$0.36 \\
2008-12-23 & 54823.463 & CH$_4$S & 20.8$\pm$0.3 & 335.6$\pm$0.3 & 1.28$\pm$0.14 \\
2010-11-29 & 55528.350 & CH$_4$S & 20.9$\pm$0.3 & 152.0$\pm$0.5 & 1.19$\pm$0.06\\
2012-01-03 & 55929.475& CH$_4$S & 23.5$\pm$0.1 & 116.3$\pm$0.1 & 1.21$\pm$0.01 \\
2012-08-12 & 56151.572 & CH$_4$S & 25.6$\pm$0.2 & 98.6$\pm$0.2 & 1.19$\pm$0.02 \\
2012-12-04 & 56265.420 & CH$_4$S & 26.7$\pm$0.1 & 91.6$\pm$0.2 & 1.23$\pm$0.01 \\
2013-08-07 & 56511.629 & CH$_4$S & 29.6$\pm$0.3 & 77.3$\pm$0.5 & 1.19$\pm$0.02 \\
2014-08-13 & 56882.584 & CH$_4$S & 34.7$\pm$0.4 & 62.2$\pm$0.5 & 1.19$\pm$0.03 \\
2014-12-09 & 57000.596 & CH$_4$S & 36.3$\pm$0.3 & 58.0$\pm$0.4 & 1.29$\pm$0.02 \\
2015-12-04 & 57360.235 & K'        	  & 41.8$\pm$0.3  & 48.4$\pm$0.4 & 1.07$\pm$0.02\\
2015-12-04 & 57360.530 & Jc        	  & 41.3$\pm$0.3  & 47.0$\pm$0.4 & 1.13$\pm$0.03\\
\hline
& & &{\bf HP~Tau/G3}&& \\
\hline
2007-11-23 & 54427.583 & K' & 29.3$\pm$1.6 & 91.8$\pm$1.4 & 1.29$\pm$0.18 \\
2008-12-21 & 54821.525& K' & 26.6$\pm$0.8 & 130.6$\pm$0.8 & 1.56$\pm$0.10 \\
2009-11-20 & 55155.478 & Kc & 22.4$\pm$0.9 & 163.1$\pm$2.1 & 1.39$\pm$0.17 \\
2010-11-28 & 55528.352 & CH$_4$S & 26.8$\pm$1.7 & 216.7$\pm$1.7 & 1.58$\pm$0.11 \\
2012-01-03 & 55929.235 & CH$_4$S & 35.7$\pm$0.2 & 244.1$\pm$0.2 & 1.60$\pm$0.01 \\
2012-08-12 & 56151.586 & CH$_4$S & 42.4$\pm$0.5 & 253.3$\pm$0.6 & 1.59$\pm$0.04 \\
2013-08-07 & 56511.626 & CH$_4$S & 52.5$\pm$0.3 & 264.7$\pm$0.2 & 1.59$\pm$0.02 \\
2014-08-13 & 56882.644 & CH$_4$S & 61.2$\pm$1.0 & 274.5$\pm$0.7 & 1.76$\pm$0.08 \\
2015-12-04 & 57360.231 & K' & 67.7$\pm$0.4 & 280.9$\pm$0.3 & 1.60$\pm$0.02\\   
2015-12-04 & 57360.543 & Jc  & 68.7$\pm$0.6 & 281.2$\pm$0.5 & 1.59$\pm$0.07\\
2018-10-31 & 58423.829 &  K' & 81.1 $\pm$0.3 &293.9$\pm$0.2 &  1.53$\pm$0.02\\
\hline
&&&{\bf Hubble~4} && \\
\hline
2007-11-23 & 54427.530 & K' & 28.4$\pm$0.1 & 106.1$\pm$0.1 & 0.40$\pm$0.01 \\
2008-12-21 & 54821.505 & K' & 17.0$\pm$0.6 & 284.3$\pm$2.8 & 1.16$\pm$0.34 \\
2008-12-23 & 54823.442 & CH$_4$S & 17.9$\pm$0.1 & 282.1$\pm$0.4 & 0.60$\pm$0.04 \\
2008-12-24 & 54823.237 & K' & 16.5$\pm$0.5 & 281.6$\pm$2.1 & 0.70$\pm$0.14 \\
2009-11-20 & 55155.493 & Kc & 40.0$\pm$0.1 & 228.1$\pm$0.1 & 0.44$\pm$0.01 \\
2010-11-29 & 55528.283 & K' & 54.4$\pm$0.1 & 207.0$\pm$0.1 & 0.36$\pm$0.01 \\
2012-01-03 & 55929.228 & CH$_4$S & 63.2$\pm$0.1 & 192.5$\pm$0.1 & 0.39$\pm$0.01 \\
2012-08-12 & 56151.542 & CH$_4$S & 65.4$\pm$0.1 & 185.9$\pm$0.1 & 0.40$\pm$0.01 \\
2012-12-02 & 56263.555 & CH$_4$S & 66.4$\pm$0.3 & 183.5$\pm$0.4 & 0.35$\pm$0.03 \\
2013-08-06 & 56510.553 & CH$_4$S & 65.7$\pm$0.1 & 175.7$\pm$0.1 & 0.37$\pm$0.01 \\
2014-08-12 & 56881.643 & CH$_4$S & 61.9$\pm$0.1 & 163.9$\pm$0.1 & 0.39$\pm$0.01 \\
2015-12-04 & 57360.239 & K' & 49.7$\pm$0.4 & 145.5$\pm$0.5      &0.34$\pm$0.03\\
2015-12-04 & 57360.263 & K' & 49.5$\pm$0.2 & 144.6$\pm$0.1    &0.39$\pm$0.01\\
2015-12-04 & 57360.525 & Jc & 50.2$\pm$0.2 & 145.2$\pm$0.4   & 0.49$\pm$0.03\\
2015-12-04 & 57360.329 & Z  & 49.0$\pm$0.6& 143.7$\pm$0.9         & 0.52$\pm$0.07\\
2016-11-08 & 57700.495 & CH$_4$S &  35.08$\pm$0.08 & 121.24$\pm$0.16 & 0.38$\pm$0.01\\
\hline
\end{tabular}
\end{table*}

\section{Taurus Binary Systems Sample} 
\label{sample}
%who discovered the visual binary nature of each object and something general about spectral type of the combined or resolved pair, flux ratio (presumably these exist at K-band?), and whether there are constraints on additional radial velocity or eclipsing companions

HP~Tau/G3 was identified as a Taurus member by \citet{cohen79} in a group of stars near HP~Tau and has an integrated light spectral type of M0.6 \citep{herczeg14}. HP~Tau/G3 was observed to be a a visual binary with contrast of $\Delta$K=1.5$\pm$0.1\,mag during the Keck non-redundant aperture masking survey of \citet{bdd2}, though it had been resolved in earlier speckle imaging (R. White, priv. comm.).  It is associated with and likely bound to HP~Tau/G2, which has a VLBI parallax of 6.2$\pm$0.03\,mas \citep{Torres09}.  HP~Tau/G3 was observed by  \emph{K2}, the repurposed \emph{Kepler} mission, in campaign 13 \citep{howell14}. Inspection of the \emph{K2} light curve rules out any eclipsing stellar companions.

FF~Tau was first identified as a Taurus star by \citet{jones79}, has an integrated light spectral type of K8 \citep{herczeg15}, and was identified as a binary system by \citet{simon87}. \citet{bdd2} measured a contrast for the visual companion of   $\Delta$K=1.04$\pm$0.02\,mag  with non-redundant aperture mask interferometry.  FF~Tau was also observed by \emph{K2}, ruling out additional eclipsing companions. Due to it's close proximity on the sky ($\sim$7\,arcmin) it is highly likely that FF~Tau belongs to the same physical association of stars as HP~Tau, and so has the same parallax as HP~Tau/G2. 

HP~Tau/G2, HP~Tau/G3 and FF~Tau likely form a gravitationally bound system, with several other objects associated with this group. Indeed, within 7 arcminutes one can find the Taurus systems HP~Tau~AB, KPNO~Tau~15, HQ~Tau, and Haro~6-28. These systems comprise $\sim$5\% of the stellar mass of the northern part of the Tau-Aur association, but only $\sim$10$^{-6}$ of the area. The alignment is therefore unlikely to be by chance, and we take all these objects to be associated. There is also no clear filamentary structure at the location of these systems in CO(1-0) maps of Taurus \citep{dame01}, adding weight to the idea that the objects are physically associated and not simply a filament seen in projection.

Investigation of the newly available Gaia DR2 \citep{gaiadr2_arxiv} parallaxes  for these systems supports this picture. With the exception of HP~Tau, all of these system have parallaxes within 1-$\sigma$ of the VLBI parallax of HP~Tau/G2 ($\omega$=6.2\,mas). HP~Tau, which appears to be in the background with a parallax of 5.65$\pm$0.11\,mas was resolved with Lunar occultation interferometry to be a binary system with separation of $<$20\,mas \citep{richichi_charm2}. The additional astrometric error term in Gaia DR2 is 0.46\,mas with a significance of 53-$\sigma$, and the astrometric renormalized unit weight error (RUWE) is 1.33 \citep{lindegren18}, implying that the Gaia DR2 parallax is not reliable with only a five parameter solution \citep{zeit8}.

Hubble~4 was first cataloged as a star thought to be associated with the reflection nebula near the highly extincted Herbig Be star V892~Tau  \citep{hubble22}, and was given a spectral type of K8.5 by \citet{herczeg15}.  Hubble~4 was identified as a visual binary with contrast of $\Delta$K=0.39$\pm$0.01\,mag with Keck non-redundant masking by \citet{bdd2}.  It is relatively bright in the radio, and was observed with VLBI to have distance of 132.8$\pm$0.5\,pc \citep{Torres07}. Hubble~4 has also been extensively monitored with spectroscopy; \citet{crockett12} identified a 0.5-1.5\,km/s RV variability on a period of $\sim$1.55\,days that was determined to be spot-driven. The presence of further spectroscopic companions is unlikely given the lack of larger amplitude RV variability. Table \ref{tprops} lists the basic properties of these three binary systems.

\section{Keck NIRC2 Observations and Analysis}
\label{keckobs}
We have monitored the orbital motion of these three binary systems over the past 12 years with the facility imager NIRC2 at the Keck II telescope, using non-redundant aperture masking interferometry (NRM) in the natural guide star AO mode. All NIRC2 AO images were taken with the smallest available pixel scale of 9.952\,mas \citep{yelda10} and the nine-hole aperture mask and multiple narrow-band IR filters. Each target was observed in one  or both of the K-band filters K' (2.124$\,\mu$m) and Kcont (2.27\,$\mu$m), the CH$_4$S filter (1.5923\,$\mu$m) and the Jcont filter (1.213,$\mu$m). Hubble~4 was also observed in the Z filter (1.0311\,$\mu$m). We employed either a two or four point dither pattern for each observation. 

The aperture masking reduction used here was the same as presented in \citet{dmys1} and \citet{bd1}, utilizing the complex triple-product, or closure-phase in addition to squared visibilities to remove non-common path errors. A binary system model, consisting of a separation, position angle and contrast, can then be fit to these observables to determine the relative astrometry and photometry at each epoch. A complete explanation of the reduction and closure-phase fitting method is given in the appendix of \citet{bd1}.  Table \ref{nirc2tab} lists the details of the observations and the fitted astrometry and magnitude differences for the three binary systems. 

\section{Orbit Fitting and Dynamical System Masses}
\label{sectorbs}

\begin{figure}
\centering
\includegraphics[width=0.4\textwidth]{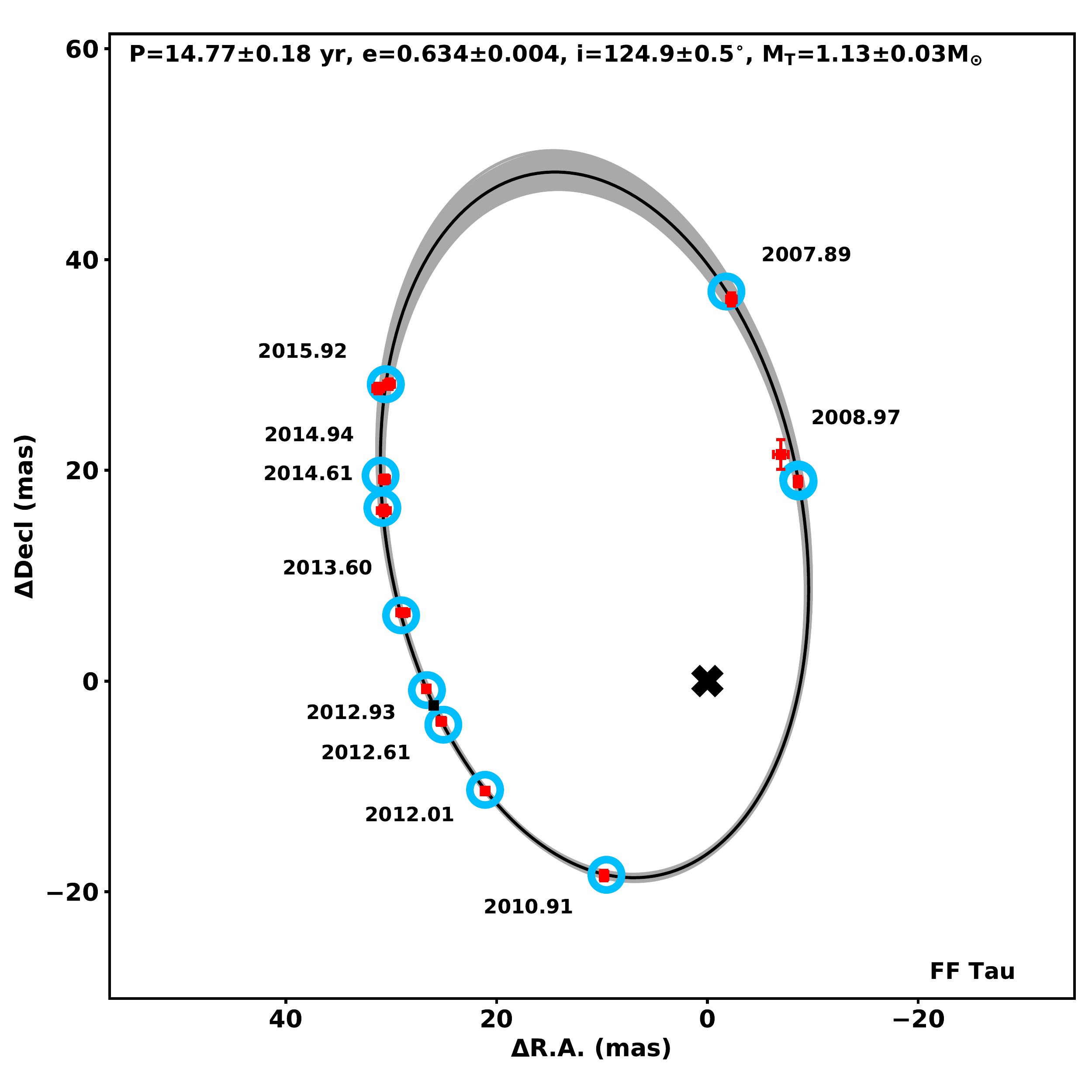}\\
\includegraphics[width=0.4\textwidth]{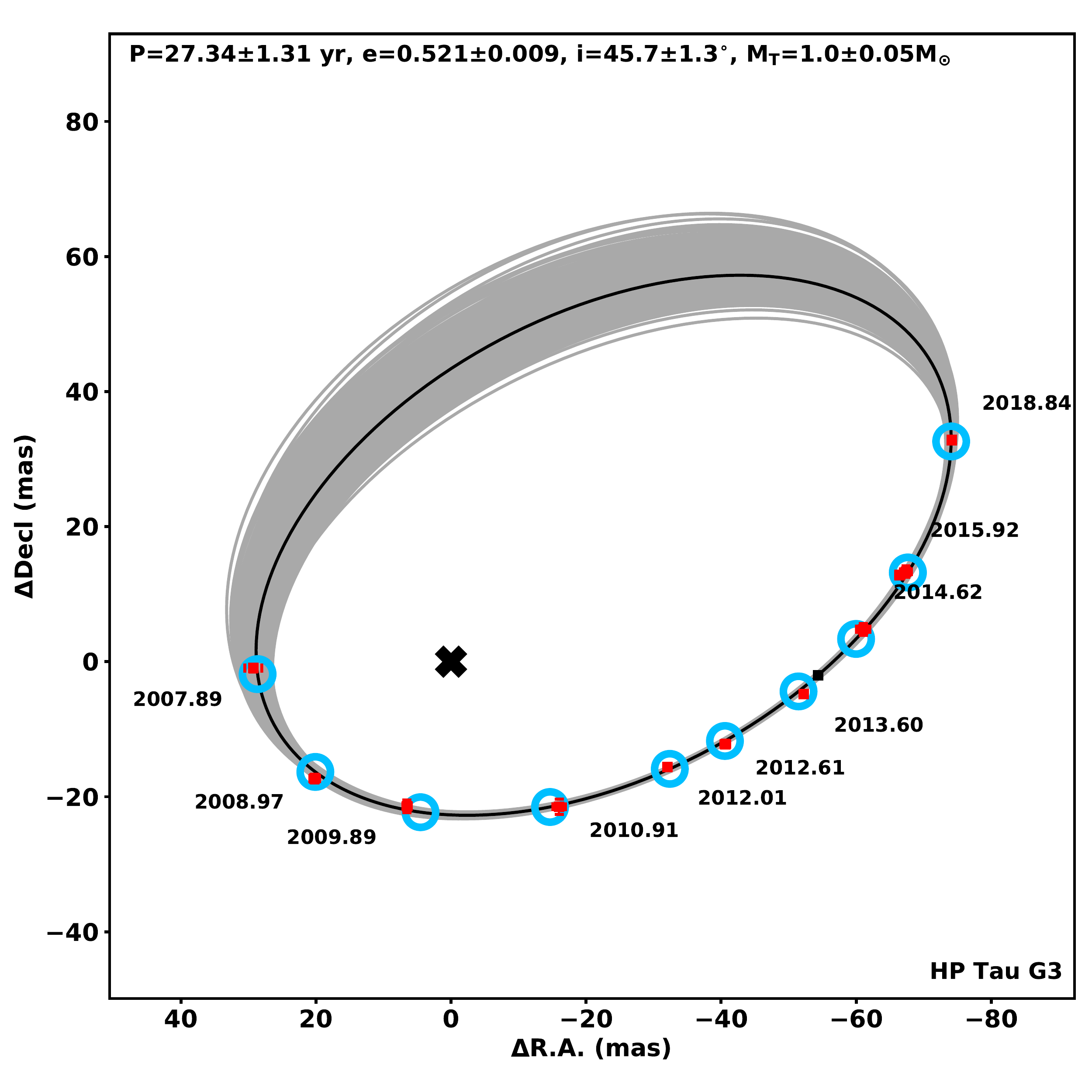}\\
\includegraphics[width=0.4\textwidth]{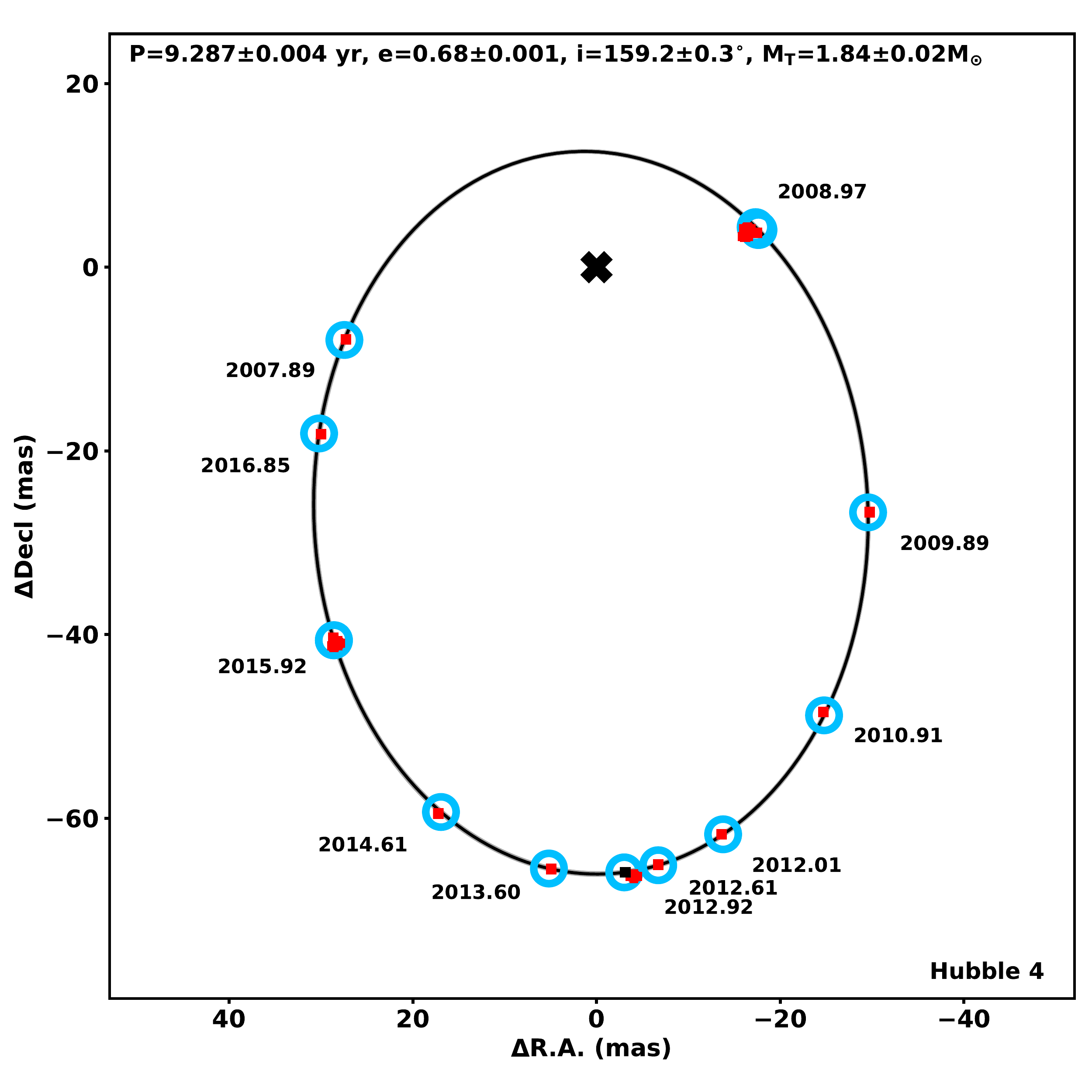}
\caption{Orbital solution for FF Tau, HP Tau/G3 and Hubble 4. The black curve is the best fit orbit, and the grey curves are 500 random orbits sampled from the fit posteriors. Red squares and blue circles indicate the observed astrometry and model predictions respectively, and the black square indicates the orbital position of the secondary at the time of observation with HST.}
\label{orbits}
\end{figure}

%%orbit table 
\begin{table*}
\caption{Orbital fits for FF~Tau, HP~Tau/G3, and Hubble~4.}
\label{orbtab}
\centering
\begin{tabular}{lccccccccc}
\hline
\hline
Name &T$_0$ & P & a& e & $\Omega$ & $\omega$ & i & M$_{\rm tot}$ &$\chi^2_r$ \\ 
&(MJD) & (days) & (mas) & & (deg) & (deg) & (deg) & (M$_\odot$) &\\
\hline
FF~Tau        & 55168.6$^{+5.7}_{-5.6}$ & 5393.5$^{+66.8}_{-63.5}$ & 38.85$^{+0.49}_{-0.47}$ & 0.634$^{+0.004}_{-0.004}$ & 176.8$^{+0.4}_{-0.4}$ & 305.2$^{+0.3}_{-0.3}$ & 124.9$^{+0.5}_{-0.5}$ & 1.129$\pm$0.027 & 3.2\\
HP~Tau/G3 & 54804$^{+28}_{-27}$ & 9984$^{+494}_{-436}$ & 56.35$^{+0.94}_{-0.76}$ & 0.521$^{+0.009}_{-0.008}$ & 292.5$^{+1.5}_{-1.7}$ & 200.8$^{+5.0}_{-4.7}$ & 45.7$^{+1.2}_{-1.3}$ & 1.005$\pm$0.053 & 3.4\\
%Hubble~4    & 54703.7$^{+0.7}_{-0.7}$ & 3394.2$^{+2.7}_{-2.7}$ & 41.75$^{+0.07}_{-0.08}$ & 0.681$^{+0.001}_{-0.001}$ & 66.1$^{+0.7}_{-0.8}$ & 68.9$^{+0.7}_{-0.7}$ & 158.7$^{+0.4}_{-0.4}$ & 1.855$\pm$0.025 & 4.4\\
Hubble~4    &54703.0$^{+0.6}_{-0.7}$ & 3392.0$^{+1.3}_{-1.3}$ & 41.64$^{+0.06}_{-0.06}$ & 0.68$^{+0.001}_{-0.001}$ & 65.7$^{+0.8}_{-0.7}$ & 68.4$^{+0.7}_{-0.7}$ & 159.2$^{+0.3}_{-0.3}$ & 1.843$\pm$0.024 & 4.4\\
\hline
\end{tabular}
\end{table*}

Orbital solutions were fit to the astrometric data for the three systems with a $\chi^2$ minimization over a grid of orbital parameters. For each system, we first generated an initial sample of 10$^4$ semimajor-axis, eccentricity, and system mass trial values, spanning 0.5-1.5 times the maximum observed orbital separation and 0.1 to 2\,M$_\odot$ in total system mass. We drew random masses rather than periods because spectral type information places useful constraints on the system masses and reduces the parameter space involved in the search. We then calculated orbital periods for each trial pair of system mass and semimajor-axis using Kepler's law. For each of the 10$^4$ random samples we then fit the remaining three orientation angles and periastron time using a Levenberg-Marquardt least squares regression. The $\chi^2$ values for the trials were then inspected in the different orbital parameters to ensure no obvious bi-modality in possible orbits was present. For all three systems, the observations spanned the majority of the full orbit and so the orbital solutions were tightly constrained and the trail semi-major axis range used was significantly larger than the region of parameter space with allowed solutions. We then further restrict the range of trial parameters and draw a new random sample for which the process is repeated. The orbit parameters with the smallest reduced $\chi^2$ value from this second sample are then taken as the starting point for a full fit over all seven orbital parameters using a Markov Chain Monte Carlo (MCMC) method.  We used \citet{emcee} implementation of the Affine-invariant MCMC, using 20 walkers initialized randomly over parameter space spanned by $\Delta\chi^2=3$ from the initial search range. We sampled the posterior 30000 times, with a 15000 step burn in and calculated the 68\% credible intervals.  Combining our orbit solutions and the literature parallax measurements, we can estimate the dynamical system masses for the three binary systems.  Table \ref{orbtab} lists the best fit orbital parameters for the three systems and Figure \ref{orbits} displays the orbital solutions.

%%Remove this figure??
%Figure \ref{orbit_corner} shows the posterior distributions for the orbital solutions.
%\begin{figure*}
%\includegraphics[width=0.5\textwidth]{FFTau_20181122_corner.pdf}
%\includegraphics[width=0.5\textwidth]{HPTauG3_20181122_corner.pdf}\\
%\centerline{\includegraphics[width=0.5\textwidth]{Hubble4_20181122_corner.pdf}}
%\caption{Posterior distributions of our orbital solutions for the three Taurus binary systems HP~Tau/G3 (\emph{left}), FF~Tau (\emph{middle}) and Hubble~4 (\emph{right}). The shading indicates the 1, 2 and 3-$\sigma$ regions.}
%\label{orbit_corner}
%\end{figure*}

\section{Hubble Space Telescope Observations}
\label{secthubble}
In addition to AO imaging with Keck/NIRC2, we have also obtained single epoch observations of these binary systems with the Hubble Space Telescope (HST) Wide Field Camera 3 (WFC3), in a variety of visible filters spanning wavelengths of 200$-$1000\,nm. Three exposures were taken in each filter, in the C512C subarray, and the standard HST reduction, calibration, and cosmic ray rejection process was applied \citep{wfc3dhb}. We then performed simple aperture photometry on the drizzled HST images with a 0.4'' radius target aperture and a sky annulus of 4$-$6'', and applied the standard WFC3 zeropoint calibration to produce unresolved magnitudes for the systems. Table \ref{hstphot} lists the unresolved system magnitudes in the WFC3 filters, and Table \ref{unrescat} lists unresolved magnitudes from 2MASS and APASS \citep{2mass,apass}. 

The epochs of the HST observations are within times spanned by the NRM orbit monitoring observations presented above, and so the relative positions of the binary components are known to $\sim$1\,mas for each system. The predicted astrometric uncertainties at the HST observation epoch are thus significantly smaller than the HST:WFC3 pixel scale ($\sim$40\,mas), and combined with the stability of the HST point-spread function (PSF) allows decomposition of the highly blended HST/WFC3 images to produce component contrast measurements in the optical  bands for the Hubble~4 and HP~Tau/G3 systems. FF~Tau was found to be too close ($<1$\,WFC3/UVIS pixel) at the time of HST observation for decomposition of the images.

Modeled after the work of \citet{garcia2015} and our previous paper \citep{dmys1}, we first assembled a library of at least 50 PSFs in each filter in the C512C subarray on the UVIS2 detector from archival data with long exposures. We visually vetted individual PSF references for elongation due to binarity, blends, or nearby cosmic rays within a few pixels of the PSF center. Other contaminants were then handled with sigma clipping in the proceeding fits. Using the Tiny Tim software \citep{tinytimproceedings} we created PSF models for each WFC3 filter and fit these to the PSF reference library to determine a modified, super-sampled PSF model that  most closely fits the library of PSF references.

\begin{figure}
\includegraphics[width=0.49\textwidth,clip,trim={9mm 0mm 9mm 9mm}]{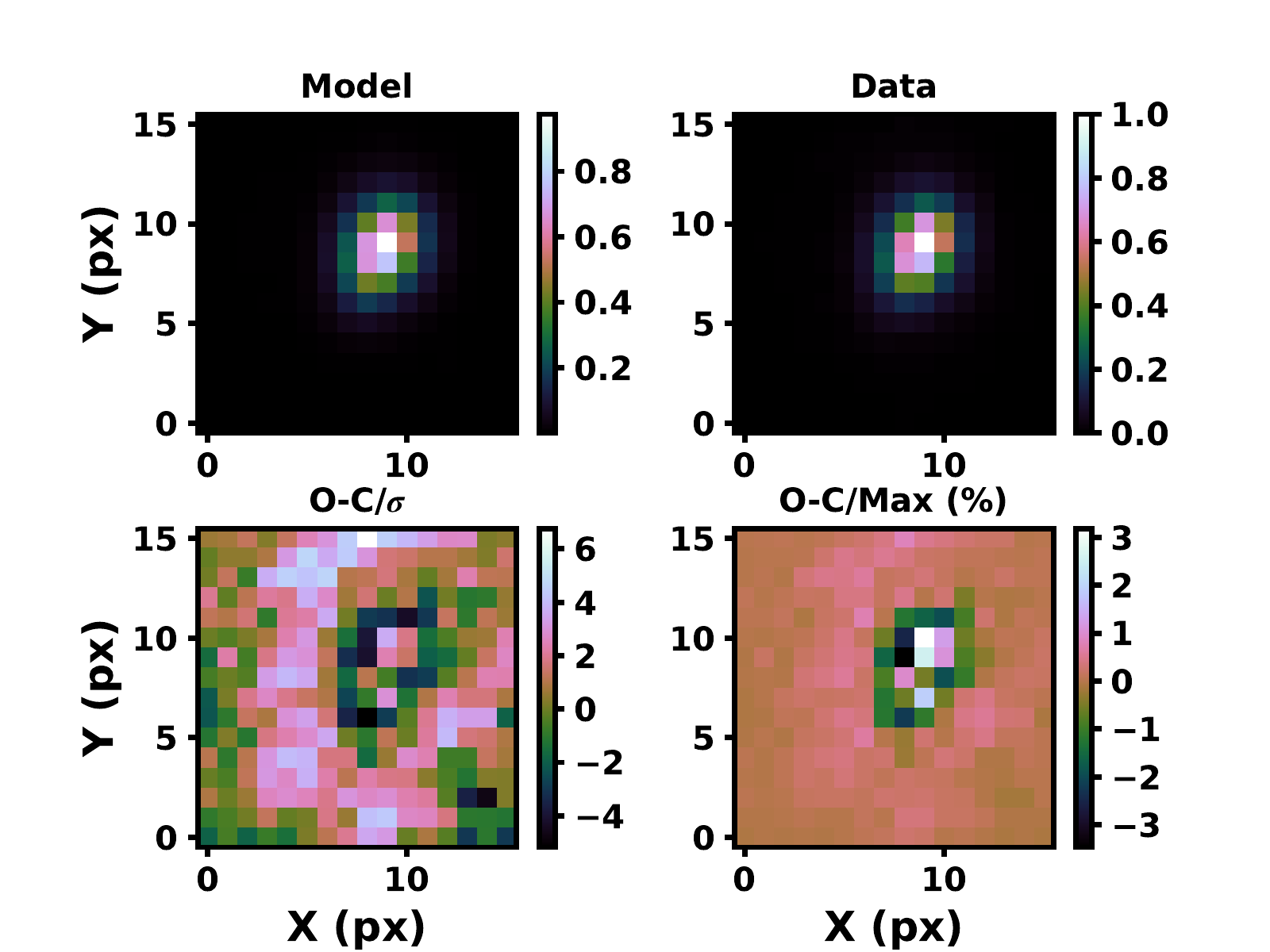}
\caption{Example two-component PSF fitting to HST:WFC3 data for Hubble~4 in the F625W filter. The model PSF consists of two sources at the separation and position angle predicted by our orbital solutions, and a 2-dimensional Gaussian blur. The resulting residuals are typically at the 1-5\% level.}
\label{hstpsf}
\end{figure}

We then fit the individual images for our binary systems using the new PSF models by sub-pixel shifting and adding the model PSF in each filter to create a model binary system with separation and position angle fixed by the orbit at  the epoch of HST observation. Because many of the HST exposures were extremely short ($<$1\,s) we expected some PSF blur induced by HST's rotational shutter. This is a well documented effect seen in exposures shorter than $\sim$5\,s \citep{hartig_shutter} and will directly affect the measured component contrasts. We model the shutter blur by applying a 2-dimensional Gaussian blur to the model PSF in the binary fitting procedure, with extent in each axis and angle allowed to vary. Figure \ref{hstpsf} displays an example fit to a single WFC3 image. Residuals each image were typically $\sim$5\% of the peak pixel value. Each object had three exposures in each WFC3 filter, and each of these images were fit separately, with different blur parameters and component contrasts. The contrasts and uncertainties were then combined with a mean to determine a final contrast ratio for the systems in each filter. Table \ref{hstrphot} lists the contrasts for the Hubble~4 and HP~Tau/G3 systems

\begin{table}
\centering
\caption{HST/WFC3 unresolved photometry.}
\label{hstphot}
\begin{tabular}{cccc}
\hline
\hline
  &  FF~Tau & HP~Tau/G3 & Hubble~4 \\
\hline
Epoch     & 2012-10-14          & 2013-12-01           & 2012-11-28         \\
F275W    & 20.38$\pm$0.39   & 20.90$\pm$0.59   & 17.61$\pm$0.10   \\
F336W    & 17.82$\pm$0.07   & 18.85$\pm$0.11   & 15.82$\pm$0.03   \\
F390W    & 16.78$\pm$0.03   & 17.72$\pm$0.04   & 15.16$\pm$0.02   \\
F395N     & 17.41$\pm$0.08   & 18.25$\pm$0.12   & 15.64$\pm$0.04   \\
F438W    & 15.91$\pm$0.03   & 16.88$\pm$0.03   & 14.46$\pm$0.02   \\
F475W    & 14.97$\pm$0.02   & 15.83$\pm$0.02   & 13.63$\pm$0.02   \\
F555W    & 14.02$\pm$0.02   & 14.87$\pm$0.02   & 12.82$\pm$0.02   \\
F625W    & 12.92$\pm$0.02   & 13.66$\pm$0.02   & 11.80$\pm$0.02   \\
F656N     & 12.06$\pm$0.03   & 12.69$\pm$0.04   & 10.84$\pm$0.03   \\
F775W    & 11.66$\pm$0.02   & 12.24$\pm$0.02   & 10.63$\pm$0.02   \\
F850LP   & 10.85$\pm$0.02   & 11.28$\pm$0.02   & 9.76$\pm$0.02    \\
\hline
\end{tabular}
\end{table}

\begin{table}
\centering
\caption{Unresolved catalog photometry.}
\label{unrescat}
\begin{tabular}{cccc}
\hline
\hline
 & FF~Tau & HP~Tau/G3 & Hubble4 \\
\hline
J & 9.78$\pm$0.02 & 10.04$\pm$0.02 & 8.56$\pm$0.02 \\
H & 8.93$\pm$0.02 & 9.15$\pm$0.02 & 7.64$\pm$0.03 \\
K & 8.59$\pm$0.02 & 8.80$\pm$0.02 & 7.29$\pm$0.02 \\
B & 15.84$\pm$0.03 &  ... &  14.35$\pm$0.08 \\
V & 13.87$\pm$0.01 &  ... &  12.69$\pm$0.05 \\
g' & 14.86$\pm$0.01 &  ... &   13.51$\pm$0.05 \\ 
r' & 13.07$\pm$0.01 &  ... &  11.96$\pm$0.04 \\ 
i' & 12.06$\pm$0.01 &  ... &  10.96$\pm$0.04 \\
\hline
\end{tabular}
\tablecomments{J, H and K magnitudes are taken from 2MASS \citep{2mass}, and the optical magnitudes are taken from APASS \citep{apass}.}
\end{table}

\begin{table}
\centering
\caption{HST/WFC3 and NIRC2 magnitude differences for Hubble~4 and HP~Tau/G3.}
\label{hstrphot}
\begin{tabular}{ccc}
\hline
\hline
Filter & Hubble~4 & HP~Tau/G3 \\
\hline
F275W & 0.78 $\pm$0.11 & ...\\
F336W & 1.02 $\pm$0.05 & ...\\
F390W & 0.96 $\pm$0.13 & 1.89$\pm$0.81\\
F395N & 0.88 $\pm$0.10 & ...\\
F438W & 0.86 $\pm$0.14 & 3.84$\pm$0.72\\
F475W & 1.00 $\pm$0.11 & .. \\ %4.3$\pm$1.3\\
F555W & 0.99 $\pm$0.25 & ...\\
F625W & 0.63 $\pm$0.12 & 3.41$\pm$0.33\\
F656N & 0.80  $\pm$0.15 & 2.20$\pm$0.28\\
F775W & 0.51 $\pm$0.26 & 2.56$\pm$0.19\\
F850LP & 0.66 $\pm$0.16 & 1.60$\pm$0.15\\
\hline
\end{tabular}
\end{table}

\begin{table}
\centering
\caption{NIRC2 NIR magnitude differences.}
\label{nirc2rphot}
\begin{tabular}{cccc}
\hline
\hline
 Filter &  FF~Tau 			& HP~Tau/G3 		& Hubble~4 \\
\hline
$\Delta$z		& ...     			& ... 				& 0.517$\pm$0.068 \\     
$\Delta$Jc		&1.132$\pm$0.033   & 1.576$\pm$0.064  & 0.486$\pm$0.032 \\
$\Delta$CH$_4$S	&1.219$\pm$0.033  & 1.601$\pm$0.028 & 0.378$\pm$0.029\\
$\Delta$K'		& 1.052$\pm$0.058  & 1.585$\pm$0.106 & 0.382$\pm$0.031\\
$\Delta$Kc		& ...				& 1.393$\pm$0.166 &  0.440$\pm$0.010\\
\hline
\end{tabular}
\end{table}

\section{Wide Field Spectrograph Observations}
\label{wifes_section}

Low-resolution spectra of the three binary systems were obtained with the Wide Field Spectrograph (WiFeS) on the Australian National University 2.3\,m telescope \citep{dopita_wifes,dopita_wifes2}. WiFeS is a dual-beam,  optical image-slicing spectrograph which provides low to mid resolution spectra over a contiguous 25'' by 38'' field of view, divided into 1$\times$0.5'' spatial pixels. In the red arm, we used the R3000 grating, which provided spectral resolution of R=3000 at wavelengths of 560-940\,nm, and in the blue arm we used the B3000 grating which provided spectral resolution of R=3000 down to 320\,nm.  Hubble~4 was observed on December 25 2015, and HP~Tau/G3 and FF~Tau were observed on 27 December 2015. The observations were taken in poor seeing, and so the data in the blue arm of the spectrograph were of low SNR ($<$10), as such we only report the red arm spectra here.

The WiFeS data were reduced using the PyWiFeS reduction packages\footnote{http://www.mso.anu.edu.au/pywifes}\citep{pywifes}. PyWiFes transforms the CCD image, consisting of a linear spectrum for each spatial pixel, into a datacube. This includes bias subtraction, flat-fielding, bad pixel and cosmic-ray removal, sky subtraction, wavelength calibration and flux calibration. The data are then interpolated to produce a consistent wavelength scale across each image pixel. We observed a flux calibrator from \citep{bessellflux99} on each of the two observing nights, which were used for flux calibration of the target spectra. This process gives a single cube for each object, with dimensions 25''$\times$38''$\times$3650 wavelength units. Following this reduction, we then applied the image combining method from \citet{wifes1_2015}, which fits a 2D-Moffat profile and background flux term to the image at each wavelength slice of the datacube and integrates the full target star flux at each wavelength to produce a linear spectrum. The resulting spectra for our targets had SNR~$=$~40$-$80 over the wavelength range.  Figure \ref{sedfit_spectra} displays the WiFeS spectra of the binary systems.

\begin{figure*}
\includegraphics[width=0.5\textwidth]{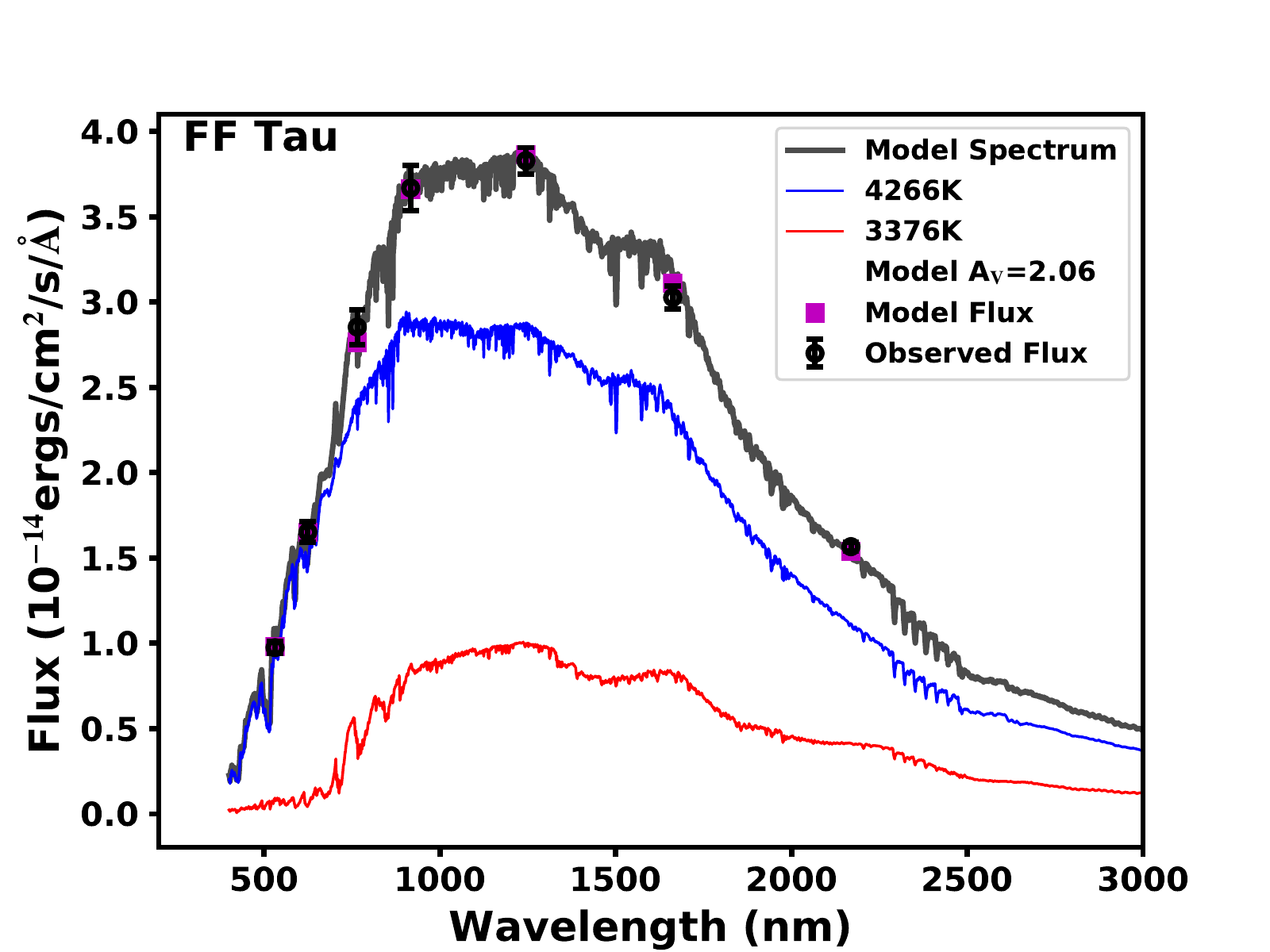}
\includegraphics[width=0.5\textwidth]{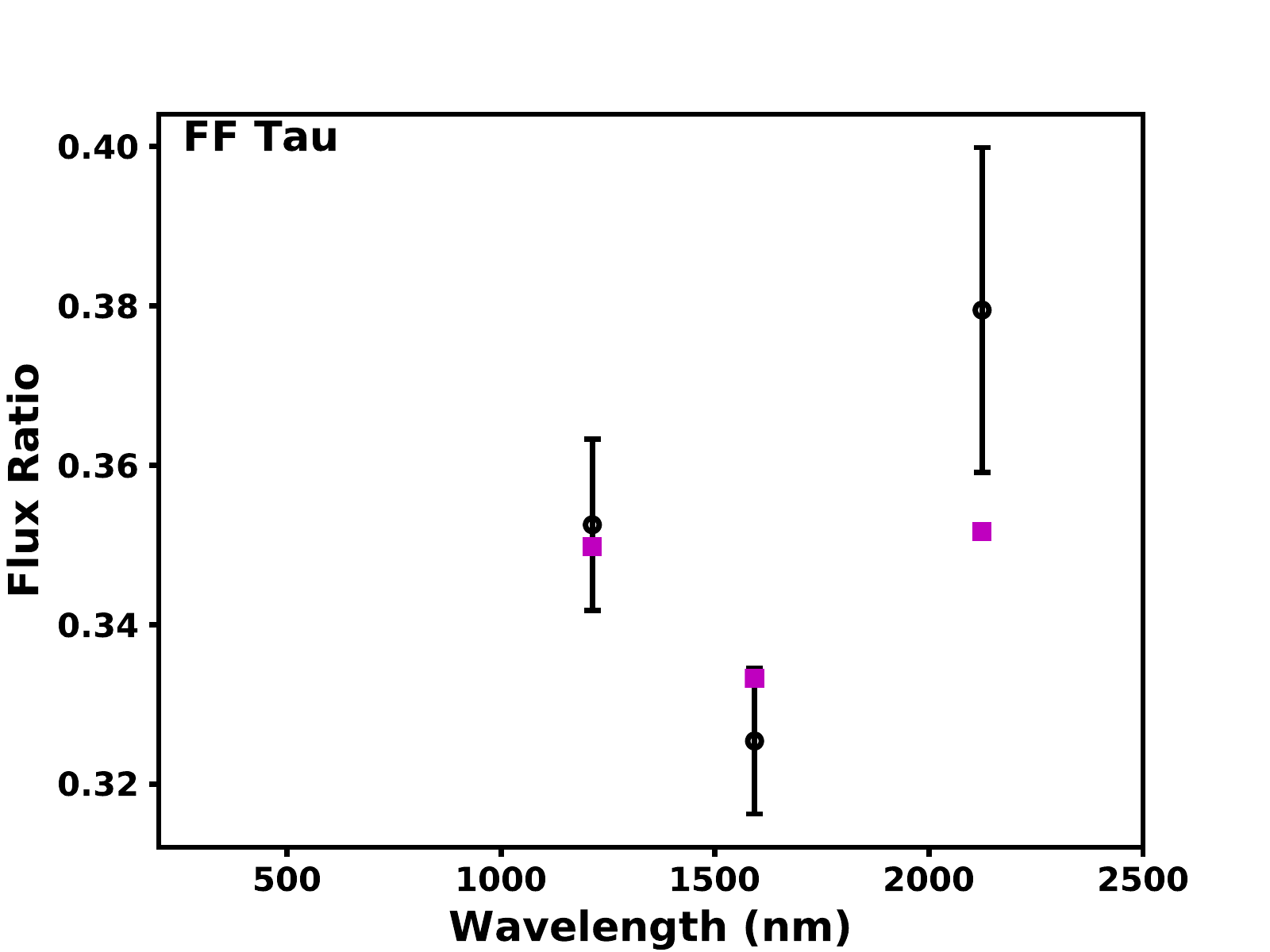}\\
\includegraphics[width=0.5\textwidth]{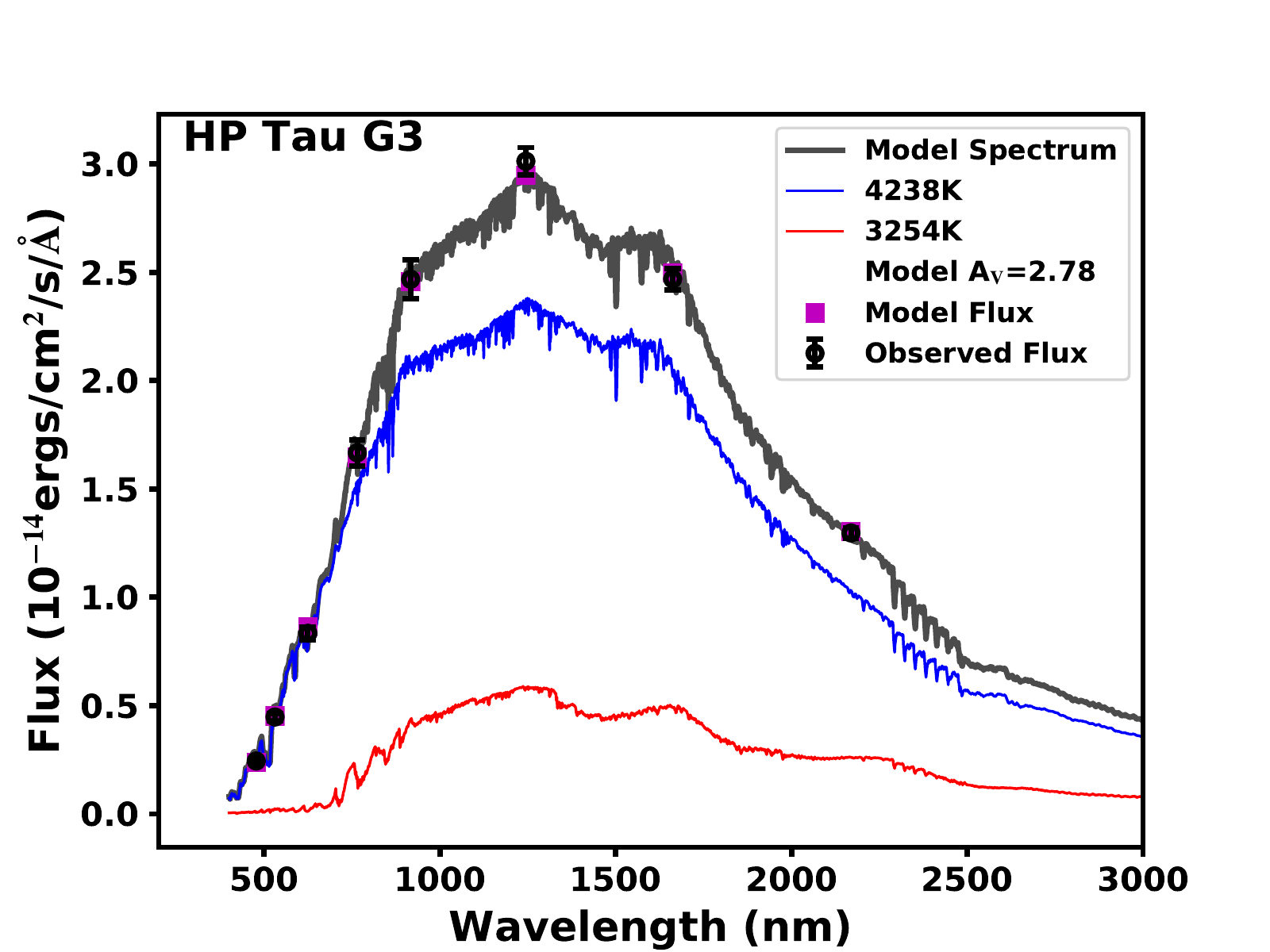}
\includegraphics[width=0.5\textwidth]{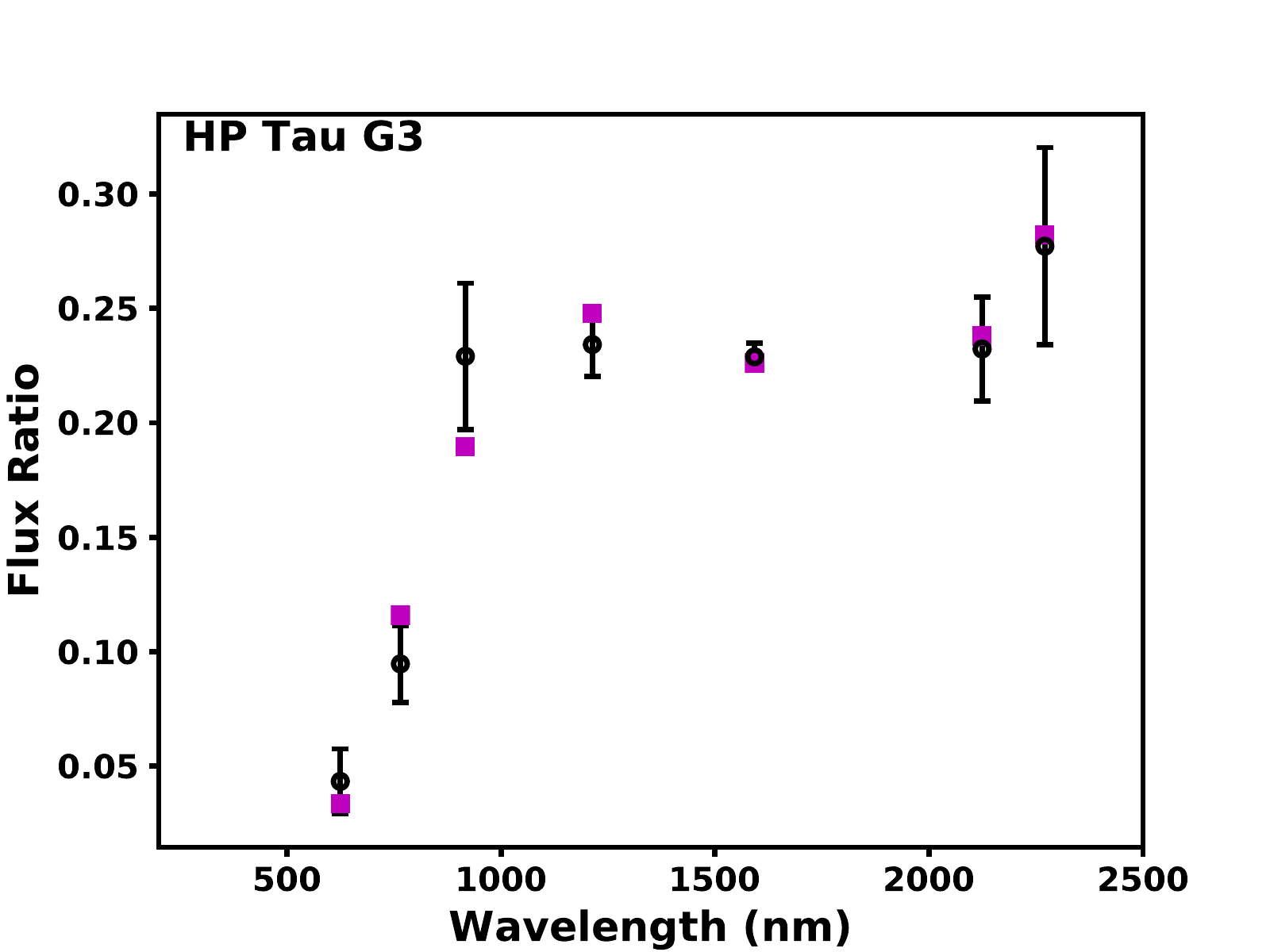}\\
\includegraphics[width=0.5\textwidth]{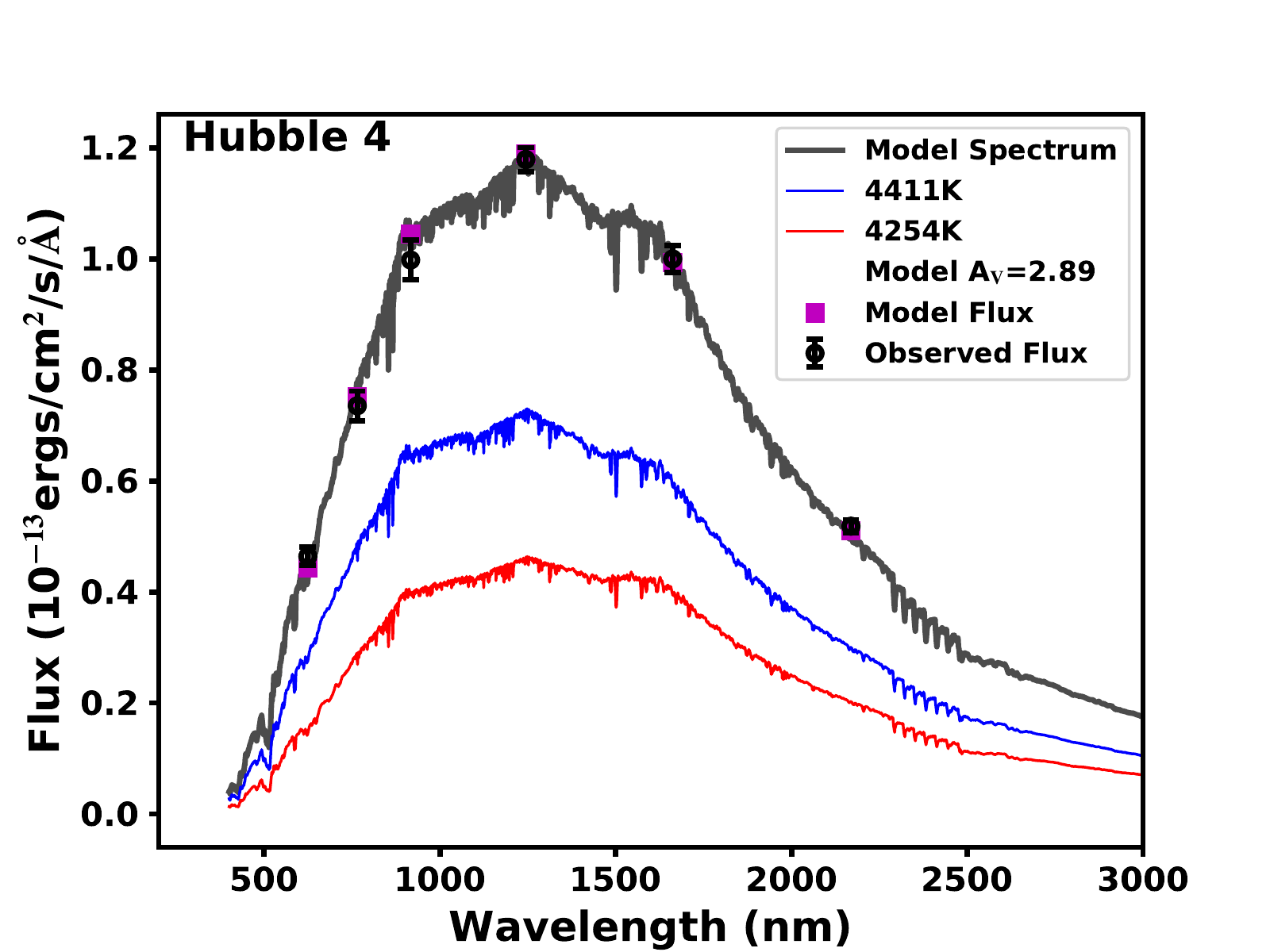}
\includegraphics[width=0.5\textwidth]{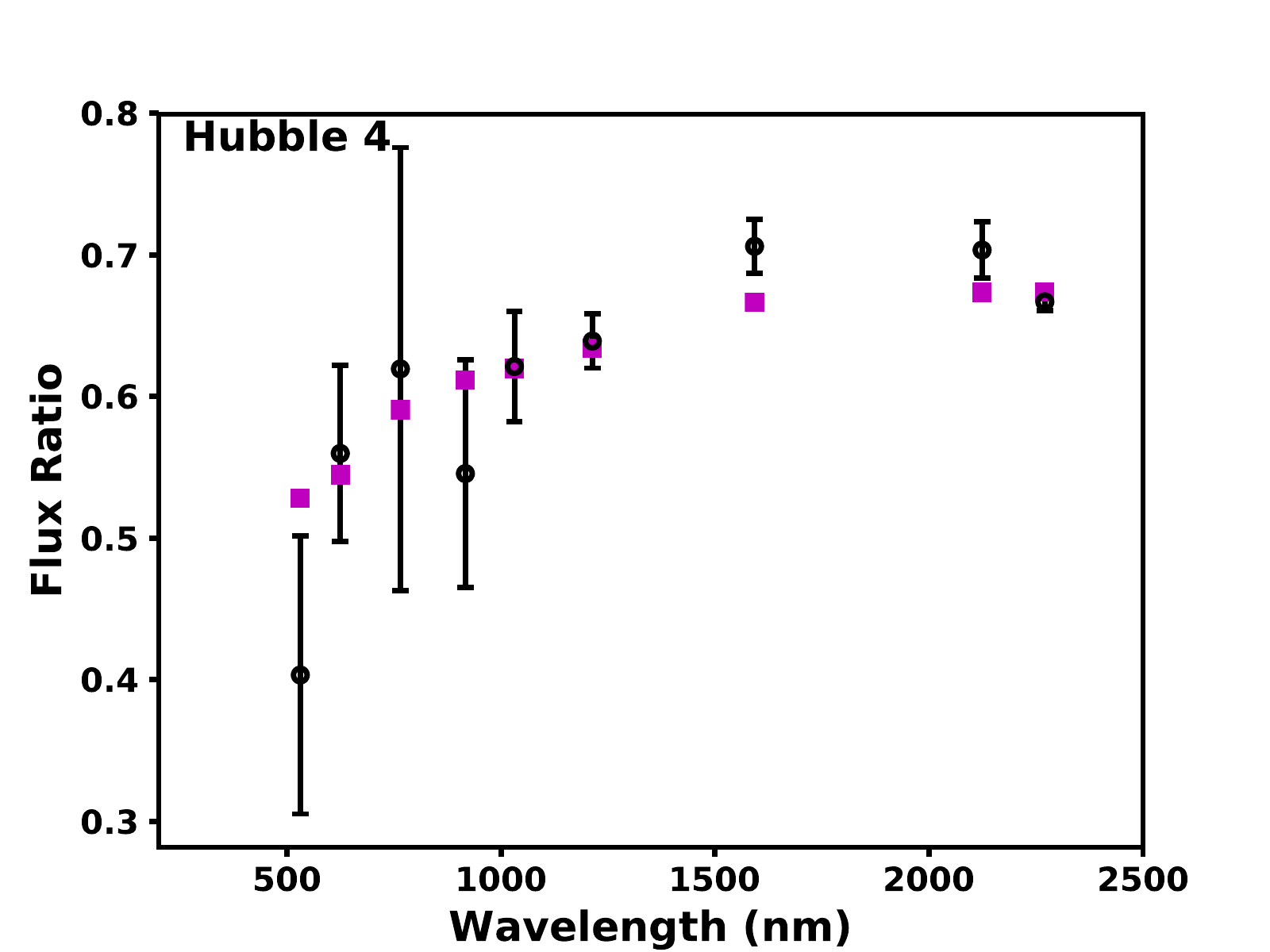}
\caption{Two component SED fits to the unresolved (\emph{left}) and resolved (\emph{right}) photometry for the three Taurus binaries in our orbit monitoring program. In the unresolved panels, the blue/red atmosphere profiles are the primary and secondary BT-Settl model atmospheres respectively, and the black profile is the combined model spectra. In all panels, black points with error bars are the measurements, and the purple squares are the forward-modeled photometry computed by integrating filter profiles on the model atmospheres and application of the best-fit extinction. Best fit model parameters can be found in Table \ref{sedfit_table}.}
\label{sedfit_plots}
\end{figure*}

\section{Two Component SED Fitting}
\label{sectSED}

Given the resolved and unresolved photometry for the the binary systems we have obtained from a combination of WFC3 imaging (Table \ref{hstrphot}), NIRC2 non-redundant masking observations (Table \ref{nirc2rphot}), and the low resolution WiFeS spectra, it is possible to decompose the combined SED and spectrum of each binary into composite profiles and fit temperatures and luminosities for the components. We use unresolved photometry from WFC3 in the optical and 2MASS in IR. We exclude other catalog photometry to avoid complication due to stellar rotation, which is common at the 1-10\% level in the optical for young ($<$100\,Myr) stars at the expected masses of these binary components \citep{zeit5}. The WFC3 data was taken in a single HST epoch, and the 2MASS IR data is significantly less contaminated by stellar variability at the longer wavelengths and as such we expect our data represents a best-case scenario for fitting SED's to variable young stars where time variability in each measurement is not expected to be significant. 

We use the BT-Settl atmosphere models in the fitting \citep{btsettl} with interpolation on the sparse grid of temperatures provided.  Following this we convolve the models with filter profiles for the WFC3, 2MASS, and NIRC2 filters of interest to produce synthetic fluxes. We then convert the measured unresolved magnitudes from the WFC3 observations to flux measurements using the appropriate zero-points for the aperture size of 0.4'' used in the aperture photometry \citep{wfc3dhb}. We then also apply Gaussian instrumental broadening of $R=3000$, and some minor rotational broadening of 20\,km/s to the component model spectra for comparison to the WiFeS spectra.

\begin{figure}
\centering
\includegraphics[width=0.49\textwidth,trim={0 0 0mm 0}]{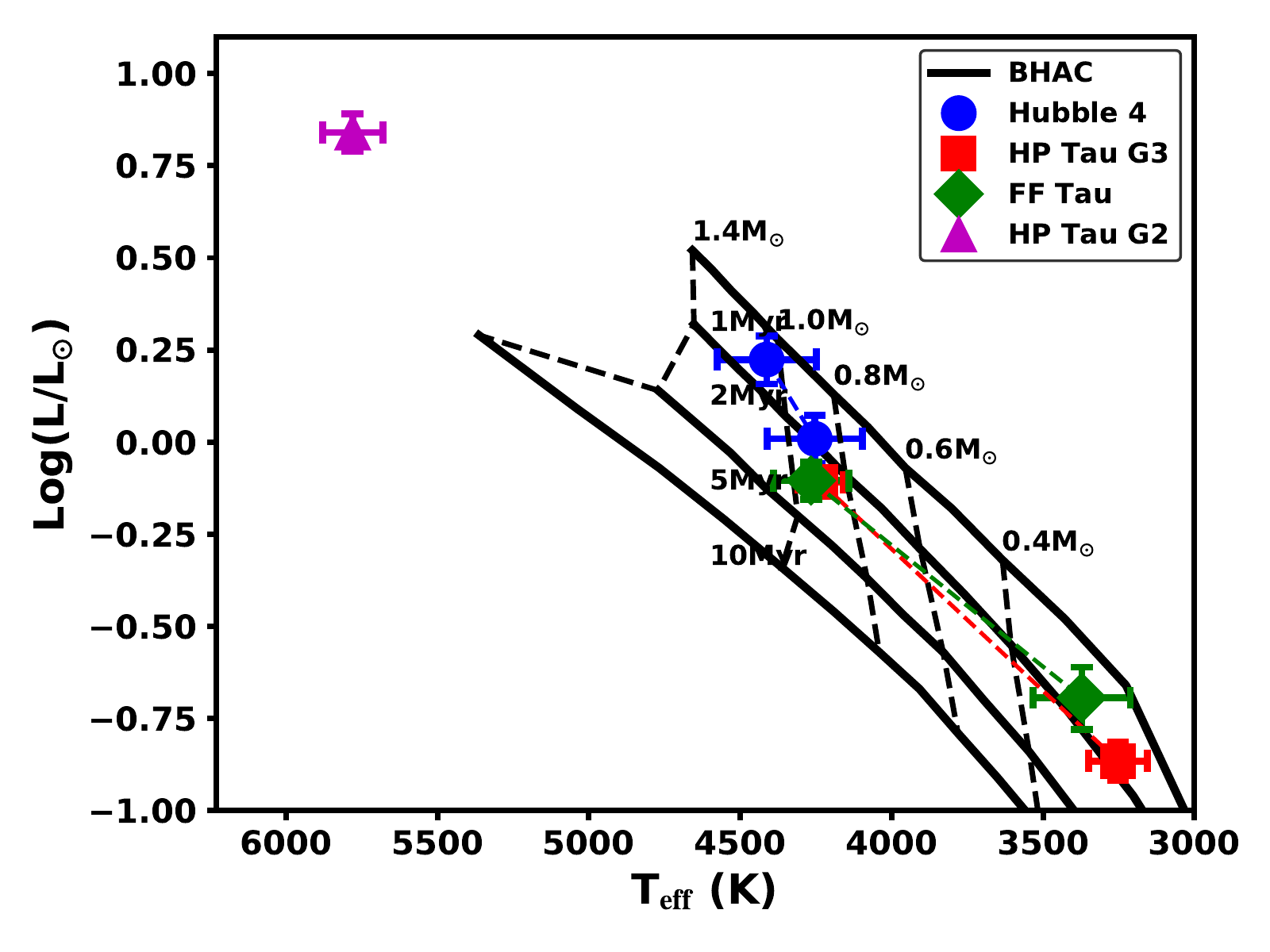}\\
\includegraphics[width=0.49\textwidth,trim={0 0 0mm 0}]{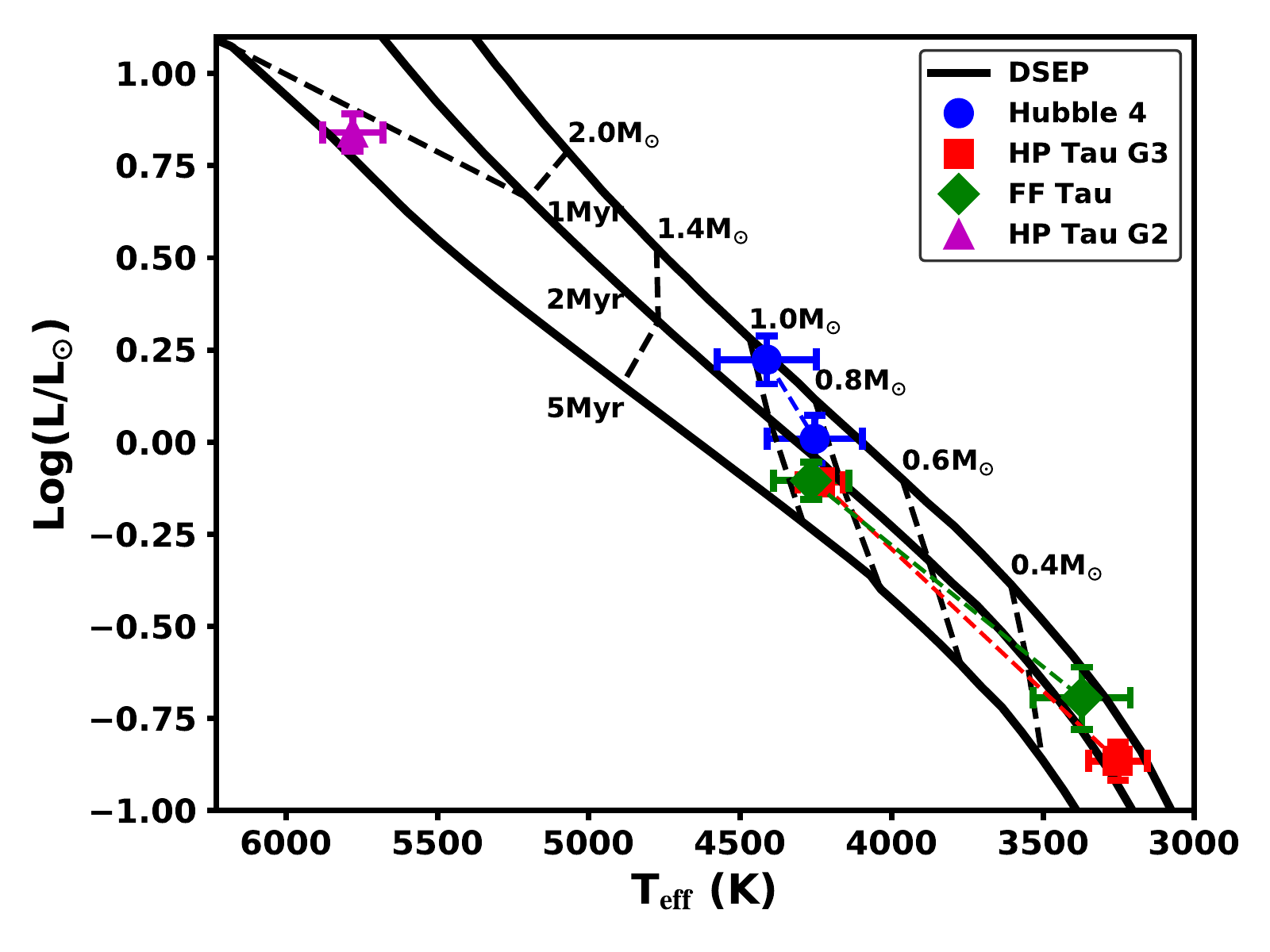}\\
\includegraphics[width=0.49\textwidth,trim={0 0 0mm 0}]{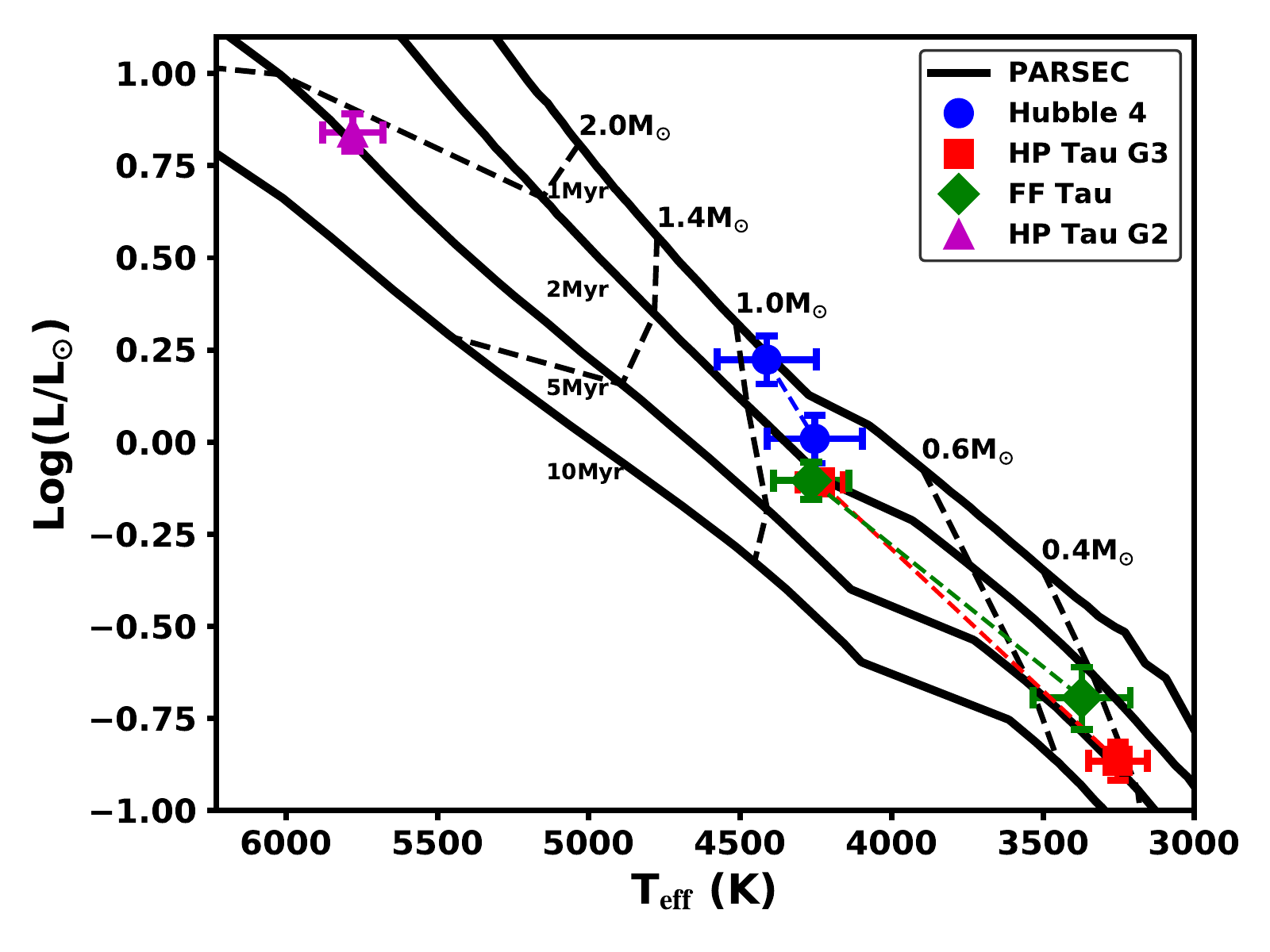}
\caption{HR-Diagram positions for the components of the three binary systems derived from the luminosity and temperature and for the three pre-main sequence models, BHAC15 \citep{bhac15} (upper) DSEP  \citep{dotter08} (middle), and PARSEC 1.2s \citep{chen14_padova} (lower). The binary system primary and secondary components are shown as colored points joined by lines. The black grid indicates the isochronal (solid) and isomass lines (dashed) for the each of the pre-main sequence models. We also show HP~Tau/G2, a single G2-type star at the same distance as HP~Tau/G3 and FF~Tau, which appears to be significantly older compared to the model grids.}
\label{modhrd}
\end{figure}

\begin{figure}
\includegraphics[width=0.5\textwidth]{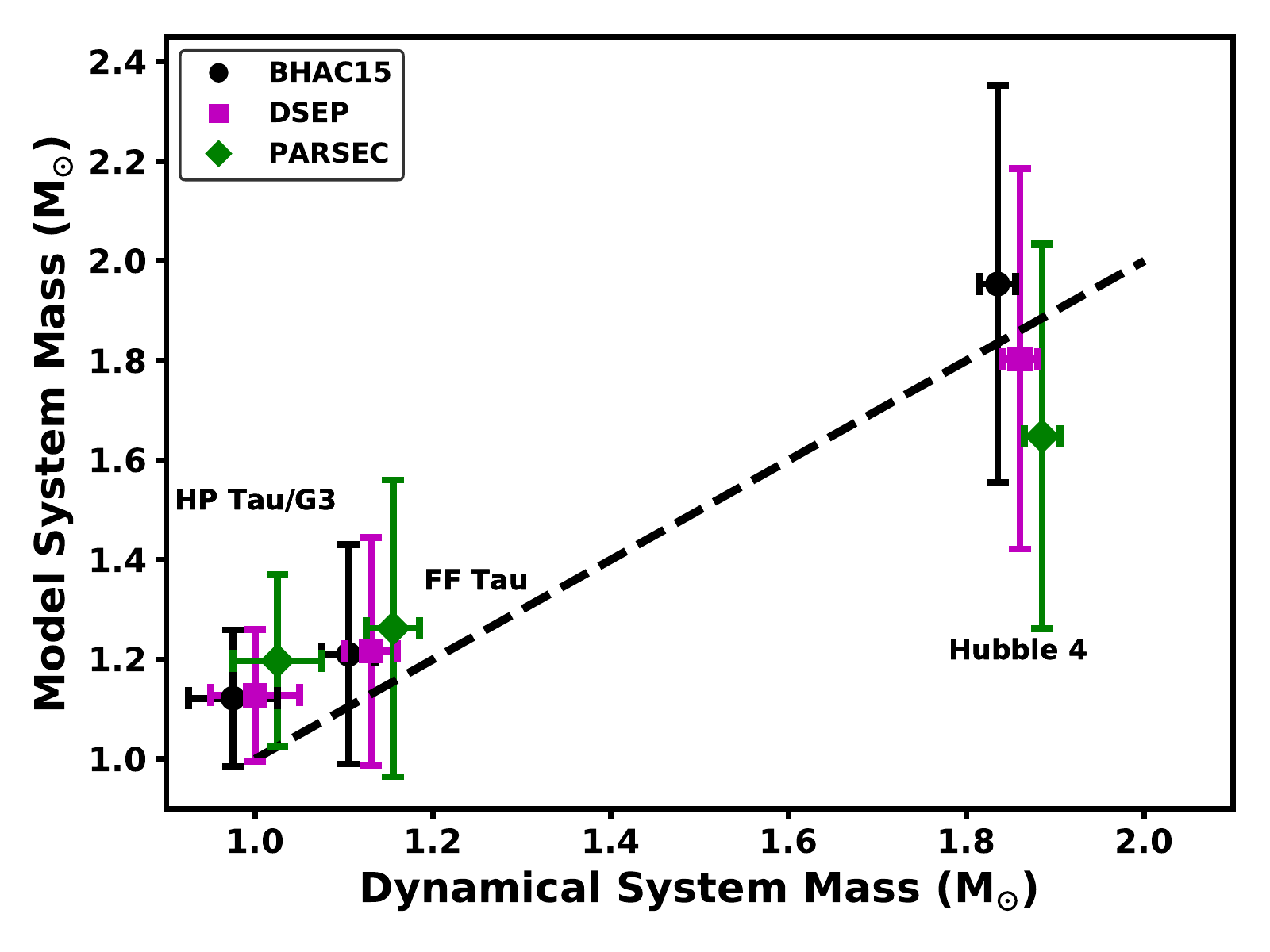}
\caption{Comparison between the dynamical system masses and masses for the components of the binary systems computed from the SED fit temperatures and luminosities for the BHAC15 \citep{bhac15} (black-circles), DSEP  \citep{dotter08} (purple-squares), and PARSEC 1.2s \citep{chen14_padova} (green-diamods) isochrones. The point are offset in dynamical mass for clarity.  There is general agreement between the models and the mass measurements with a slight systematic offset to higher model temperatures for HP~Tau/G3 and FF~Tau. The SED fit temperature and luminosity uncertainties dominate the error budget, mainly due to the lack of precision resolved photometry in the optical.}
\label{tmassplot}
\end{figure}

We fit a six component model to the resolved and unresolved photometry, consisting of two model temperatures, a radius ratio term, a reddening parameter, and an overall flux scale for both the photometry and the low-resolution spectrum. For the reddening, we interpolate the \citet{savage_mathis79} reddening law to each filter and apply it to the model photometry and component model spectra.  We initially try a small grid of primary and secondary temperatures with starting points chosen based on the integrated light spectral types and IR flux ratios, and then take the best grid-point as starting parameters for a Levenberg-Marquardt least squares regression. We deliberately exclude any photometry blue-ward of the F555W filter, as the shorter wavelength filters are typically poorly fit by models for young stars. We also apply a 3-$\sigma$-clip to reject any additional photometry that is poorly fit by the models, this resulted in rejection of the F555W unresolved photometry for Hubble~4. 

Figure \ref{sedfit_plots} shows the final SED fits to the data for the three binary systems, including both the primary and the secondary component contributions to the total flux at each wavelength, and Figure \ref{sedfit_spectra} show the model comparison to the WiFeS spectra.  We then determine the component luminosities by integrating the model atmosphere fluxes at the best-fit temperatures according to the flux-scale and ratio terms, and scaling by distance. Table \ref{sedfit_table} lists the best-fit temperatures, luminosities, and corresponding radii. We were unable to produce a two component model that fit both the spectrum and the photometry for the Hubble~4 system: While the photometry alone is able to be fit with two components of temperature $T_{\mathrm{eff}}>4100$\,K, the WiFeS spectrum shows the characteristic TiO regions of a much lower temperature primary. We discuss this further below.

Following the SED fitting, we then compare the component temperatures and luminosities to evolutionary tracks from the  BHAC models \citep{bhac15} to determine component ages and masses to compare to the total system masses from derived from the orbits.  Figure \ref{modhrd} shows the HR-diagram positions of the components of the three binary systems in relation to the 1-10\,Myr BHAC \citep{bhac15}, DSEP \citep{dotter08} and PARSEC 1.2s \citep{chen14_padova} isochrones and the corresponding total system masses compared to the dynamical masses. For all three models, the total system mass of HP~Tau/G3 derived from the models is $\sim$1-$\sigma$ offset from the empirical values towards larger masses. FF~Tau is also offset to higher model masses but by a smaller margin. Figure \ref{tmassplot} shows the comparison of the dynamical system masses to the model-derived total system masses.

\section{Model Comparison}
\label{sectmodmod}
%%!!!!REWRITE THIS APPARENTLY TO BE CLEARER
The spectral types for these three binary systems were measured to be K7 \citep{kenyon95}, and more recently updated to K8-M0.5 using optical spectra \citep{herczeg14}. These spectral types imply somewhat cooler effective temperatures than what we find in the SED fitting. Veiling from accretion in the optical is unlikely to have introduced a significant spectral slope, as these stars do not have observable disk material. Indeed, \citet{herczeg15} estimated that the effect of veiling in the optical for these systems was negligible. For the cases of FF~Tau and HP~Tau/G3, we expect that the  combined light spectra, variable extinction in the Taurus clouds, and the steep age-mass gradient at this point on the pre-main sequence is the likely cause for the small difference in  integrated-light spectral types and our two component SED temperatures. For the Hubble~4 system, we discuss below that the data is most readily explained by the presence of a third, as yet unresolved, component to the system.
%%The model masses  for these systems are roughly proportional to the estimated extinction parameters, i.e., the lower mass primaries have smaller values of extinction, which is expected given the integrated light spectral types.

\begin{figure}
\includegraphics[width=0.5\textwidth]{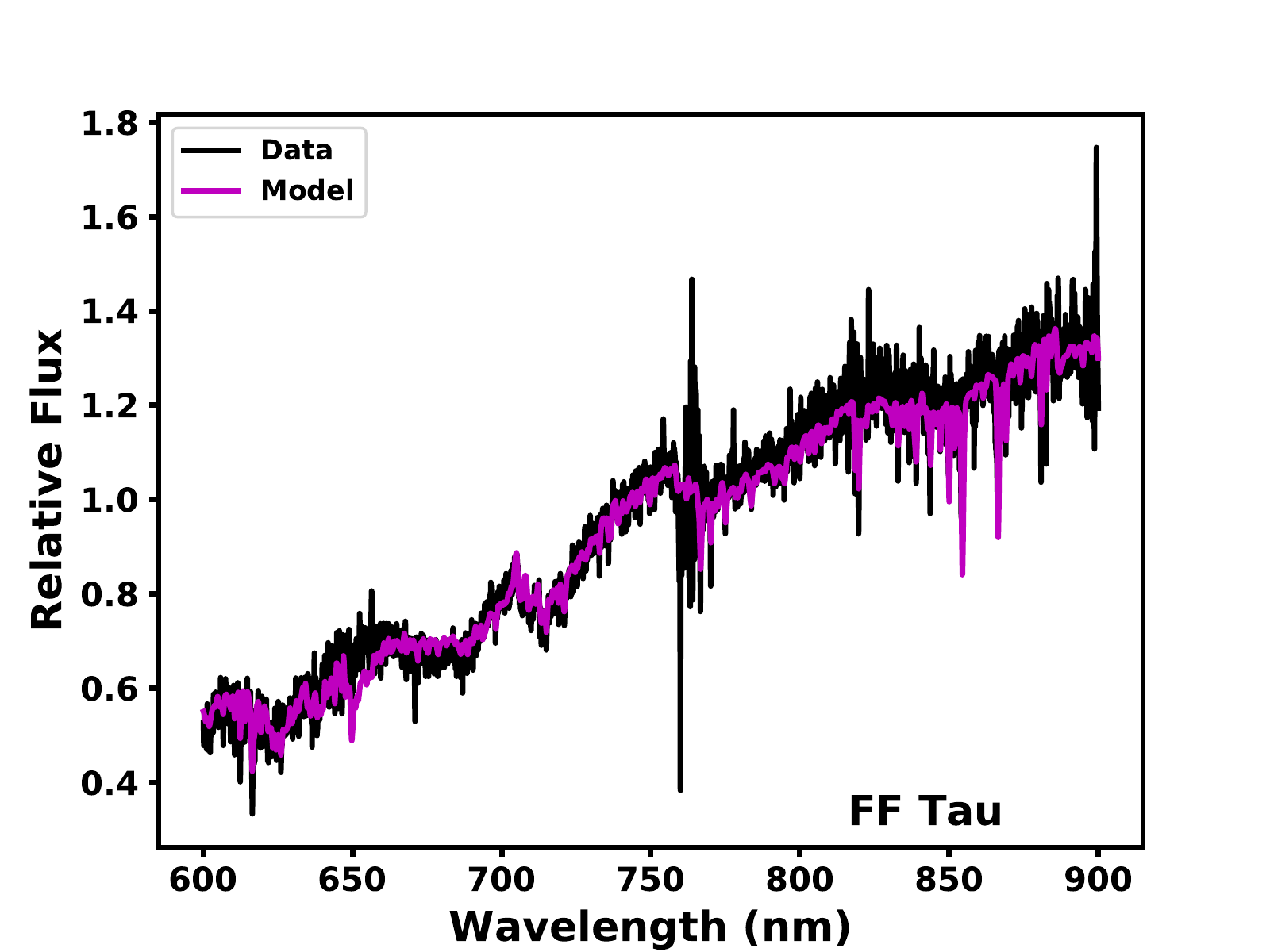}
\includegraphics[width=0.5\textwidth]{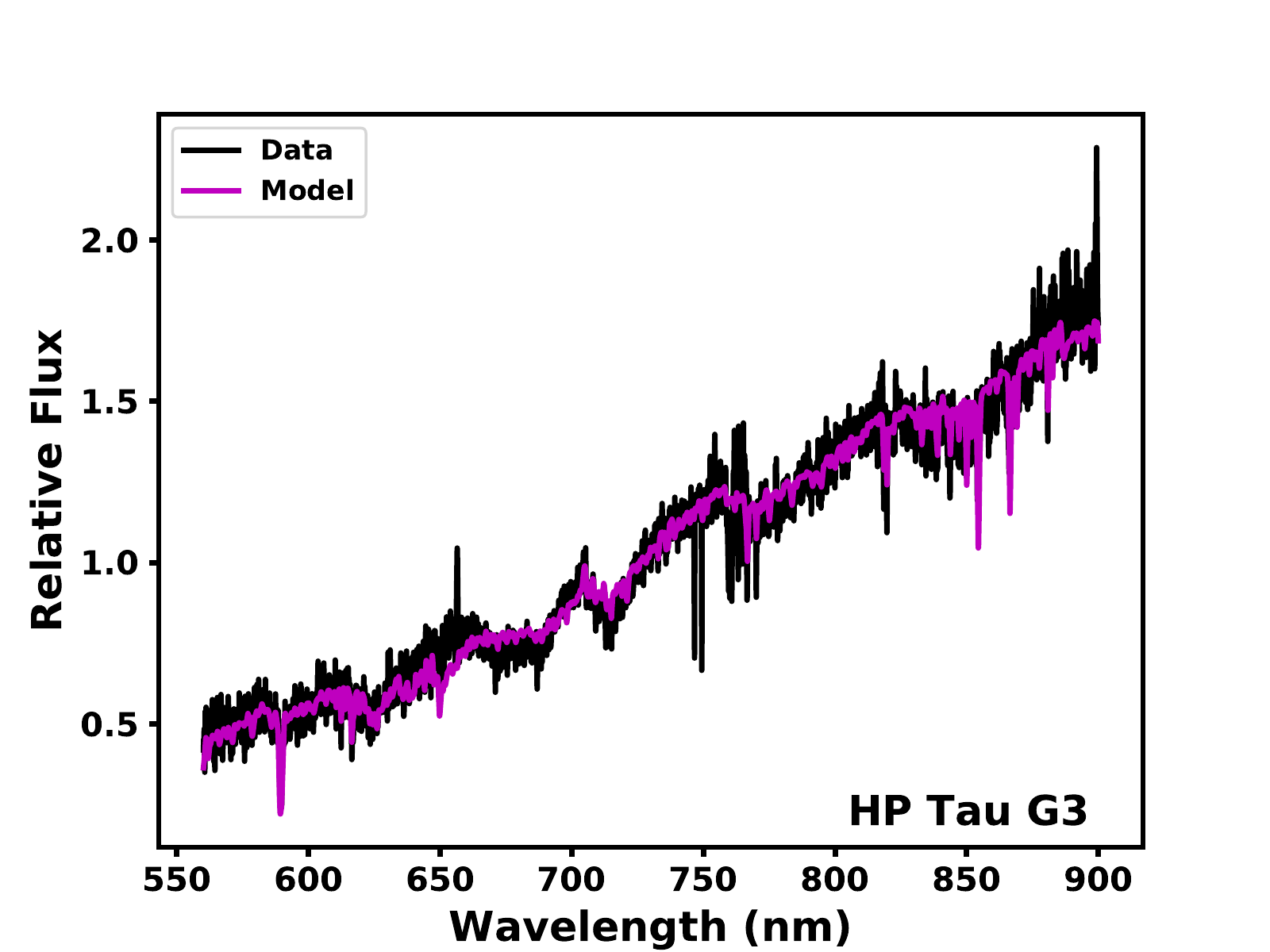}
\includegraphics[width=0.5\textwidth]{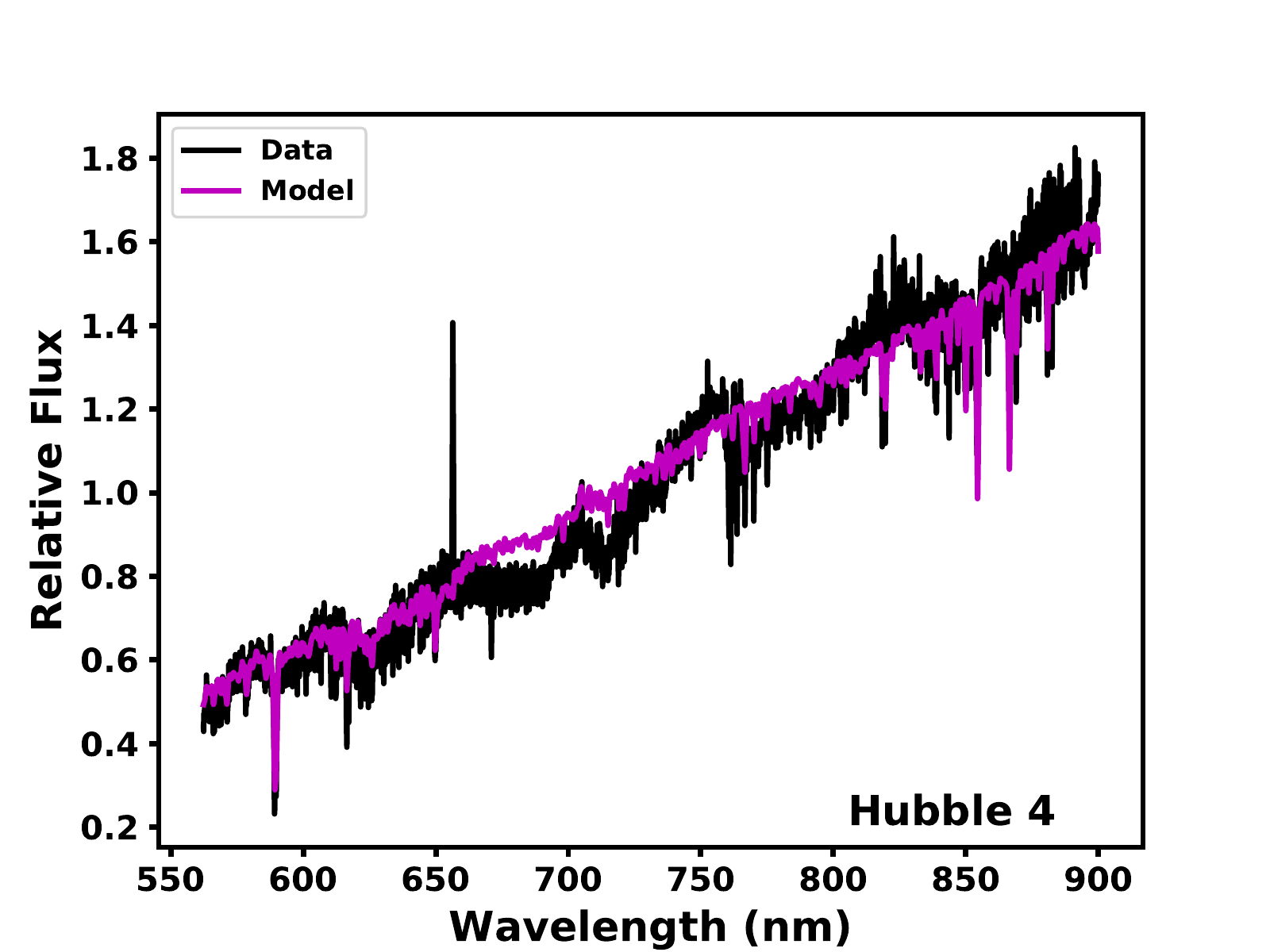}
\caption{The sum of the two component model spectra from the SED fitting compared to unresolved WiFeS spectra. The spectra for FF~Tau and HP~Tau/G3 match the SED fit profile relatively well, despite expected differences due to the youth of the sources. The unresolved spectrum of Hubble~4 is significantly different to the SED fit component temperature combined spectrum. Given the NIR flux ratios observed in the orbit monitoring data and the component masses from \citep{galli18}, we suggest that Hubble~4 may be a hierarchical  triple system, with Hubble~4 A being an as yet unresolved binary system.}
\label{sedfit_spectra}
\end{figure}

We also compared the best fit two component model spectrum for each binary system to the unresolved WiFeS spectra. The WiFeS spectra and model SED fit spectra are shown in Figure \ref{sedfit_spectra}. The spectra for FF~Tau and HP~Tau/G3 both qualitatively match the two-component SED model in the 560-900\,nm wavelength range, indicating the  temperatures and reddening terms we infer from the resolved and unresolved photometry are consistent (Figure \ref{sedfit_spectra}). The observed spectrum for Hubble~4 is significantly different from the SED model. 

\begin{deluxetable}{lccc}
\tabletypesize{\scriptsize}
\tablecaption{SED fit component temperatures, luminosities, radii, and reddening, and corresponding model parameters for the components of the three Taurus binary systems. Note that the parameters for Hubble~4 assume a two component fit and ignores the discrepancy with the low-res spectra (see above). \label{sedfit_table}}
\tablehead{\colhead{} & \colhead{FF~Tau} & \colhead{HP~Tau/G3} & \colhead{Hubble~4}}
\startdata
T$_{\mathrm{eff},p}$ (K)	& 4266$\pm$124	& 4238$\pm$75	& 4411$\pm$164 \\
T$_{\mathrm{eff},s}$	(K)	& 3376$\pm$160	& 3254$\pm$100	& 4254$\pm$156  \\
L$_p$ (L$_\odot$)		& 0.79$\pm$0.10	& 0.78$\pm$0.06	& 1.67$\pm$0.26 \\
L$_s$ (L$_\odot$)		& 0.20$\pm$0.04 	& 0.14$\pm$0.02	& 1.02$\pm$0.16\\
R$_p$ (R$_\odot$)		& 1.62$\pm$0.14	& 1.64$\pm$0.08 	& 2.21$\pm$0.24 \\
R$_s$ (R$_\odot$)		& 1.32$\pm$0.20	& 1.16$\pm$0.10	& 1.86$\pm$0.20 \\
E(B-V) (mag)			& 0.66$\pm$0.13	& 0.89$\pm$0.05	&  0.93$\pm$0.10 \\
\hline
\multicolumn{4}{c}{BHAC Fit Parameters}\\
\hline
Age$_p$ (Myr)			& 3.4$\pm$1.3	        & 3.2$\pm$0.8	        & 1.5$\pm$0.8 \\
Age$_s$ (Myr)			& 1.6$\pm$0.7 	        & 1.9$\pm$0.8  	        & 1.9$\pm$0.8 \\
M$_p$ (M$_\odot$)		& 0.94$\pm$0.14	& 0.9$\pm$0.1 	        & 1.06$\pm$0.2 \\
M$_s$ (M$_\odot$)		& 0.27$\pm$0.08	& 0.22$\pm$0.05	& 0.89$\pm$0.2 \\
M$_\mathrm{Tot}$ (M$_\odot$)	& 1.21$\pm$0.22	& 1.12$\pm$0.14	& 1.95$\pm$0.4 \\
\hline
\multicolumn{4}{c}{DSEP Fit Parameters}\\
\hline
Age$_p$ (Myr)			& 3.1$\pm$1.5	        & 2.9$\pm$0.9        & 1.1$\pm$0.8 \\
Age$_s$ (Myr)			& 1.5$\pm$1.1 	        & 1.7$\pm$0.7 	        & 1.7$\pm$0.8 \\
M$_p$ (M$_\odot$)		& 0.92$\pm$0.14	& 0.9$\pm$0.1 	        & 0.95$\pm$0.20 \\
M$_s$ (M$_\odot$)		& 0.29$\pm$0.09	& 0.23$\pm$0.05	& 0.85$\pm$0.18 \\
M$_\mathrm{Tot}$ (M$_\odot$)	 & 1.22$\pm$0.23	& 1.13$\pm$0.13	& 1.8$\pm$0.4 \\
\hline
\multicolumn{4}{c}{PARSEC Fit Parameters}\\
\hline
Age$_p$ (Myr)			& 2.5$\pm$1.5	        & 2.2$\pm$0.9          & 1.1$\pm$0.8 \\
Age$_s$	(Myr)			& 3.5$\pm$2.5         & 4.7$\pm$1.5 	        & 1.5$\pm$0.8 \\
M$_p$ (M$_\odot$)	        & 0.80$\pm$0.15	& 0.77$\pm$0.08      & 0.88$\pm$0.21 \\
M$_s$ (M$_\odot$)		& 0.46$\pm$0.15	& 0.43$\pm$0.09	& 0.76$\pm$0.17 \\
M$_\mathrm{Tot}$ (M$_\odot$) & 1.26$\pm$0.30	& 1.20$\pm$0.17	& 1.65$\pm$0.38\\
\enddata
\end{deluxetable}

\subsection{Hubble~4}
The TiO bands present in the WiFeS spectrum (Figure \ref{sedfit_spectra}) indicate a cooler temperature for the components than what we infer from the unresolved photometry and magnitude differences. We note that the WiFeS spectrum implies a combined light spectral type of K7-M0, which is consistent with the temperature estimate of 3900\,K from \citet{herczeg15} and is significantly cooler than either of the components we fit to the Hubble~4 SED (Figure \ref{sedfit_plots}). 

The discrepancy remains for Hubble~4  when considering the dynamical masses: It is difficult to reconcile the spectral type from the optical spectra with the  dynamical system mass of 1.843$\pm$0.024\,M$_\odot$ and the NIR secondary-to-primary flux ratio of $\sim$0.65. \citet{galli18} measured the component masses of the Hubble~4 system using Very Long Baseline Interferometry (VLBI), in combination with the orbit presented in this paper, and found the components to be 1.234$\pm$0.023\,M$_\odot$ and 0.730$\pm$0.020\,M$_\odot$ respectively. A 1.23\,M$_\odot$ star at $<$5\,Myr is expected to be significantly hotter that 3900\,K according to multiple stellar evolution models \citep{bhac15,dotter08,chen14_padova}. We suggest that Hubble~4 may be a hierarchical triple system, with the primary consisting of two stars of unequal mass, with the effective temperature of the more massive component closer to $\sim$4000\,K.

Given the extent and variety of the observations of the  Hubble~4, there is only a small region of parameter space in which a third component to the system could exist. Inside the orbit of the known companion, observations using VLBI  rule out additional stellar companions at angular separations $>$3\,mas \citep{galli18}, and radial velocity monitoring rules out spectroscopic companions \citep{crockett12}. The only remaining possible configuration is a near face-on orbit ($i\simeq0$) with separation $<$0.5\,AU. The secondary to primary mass ratio of the unresolved components must also be smaller than unity to produce the size and slope the Hubble~4\,A-B optical and NIR flux ratios. Such a companion may be detectable with additional VLBI monitoring if the posited component exhibits radio emission. Such a component may produce a orbital radial velocity signal smaller than the  rotational variability-produced signal from Hubble~4\,B,  which would be expected to dominate if the two components of Hubble~4A are pole-on towards Earth.

\subsection{HP~Tau/G3 and FF~Tau}
For the two systems with mid-M secondaries, HP~Tau/G3 and FF~Tau, the PARSEC  isochrones in both cases produce older ages by up to $\sim$2\,Myr for the secondary components. The PARSEC evolutionary models employ the  PHOENIX BT-Settl model atmospheres \citep{btsettl} for stars cooler than 4700\,K, to produce the synthetic color-temperature/optical depth relations \citep{chen14_padova}. This is then adjusted empirically to better match the colors of  M-dwarf members of intermediate age clusters Praesepe and M67. It is unclear how this calibration to older M-dwarfs might affect our model fitting in the pre-main sequence for cooler stars. We expect that the systematic differences between the PARSEC models and the other two grids is produced by the calibration methodology.

The FF~Tau and  HP~Tau/G3 primary components both have model derived ages of $\sim2.5-3.5$\,Myr, which is within the expected range for K-type stars in the Taurus clouds \citep{taueco1}. This age is in significant disagreement with the age of the nearby star HP~Tau/G2. HP~Tau/G2 is at the same distance and is associated with the both FF~Tau and HP~Tau/G3, and is likely bound to the latter. \citet{bdd2} surveyed HP~Tau/G2 with NIRC2 coronagraphy and aperture masking, and did not find a nearby companion.   HP~Tau/G2 has a spectral type estimated from optical spectra of G2 \citep{herczeg14}, which corresponds to a temperature of 5690\,K according to their temperature scale and  Log$(L/L_\odot)=0.84\pm0.10$ at the measured distance of the system of 161\,pc \citep{Torres09}. The \citet{pecaut13} spectra-type to temperature conversion gives an effective temperature of 5870\,K which is in agreement with \citet{herczeg14} within the uncertainties of the temperature scales. These values place HP~Tau/G2 at a position on the HR-diagram corresponding to an age closer to $\sim$5\,Myr, (Figure \ref{modhrd}) which is significantly older than the mean age of the three lower-mass binaries ($\sim$2.5\,Myr). This mass-age trend in the models extends to the companions to FF~Tau and HP~Tau/G3.  In the comparison to both the DSEP and BHAC15 models, the model ages for the secondary components determined from temperatures and luminosities are systematically younger than the primaries by a factor of two.

The stellar membership of the Taurus-Auriga star-forming region is certainly not a coeval population. There are clear regions of ongoing stars formation surrounded by $\sim1-3$\,Myr old pre-main sequence stars \citep{luhman09}, with spatial and kinematic subclustering \citep{galli_taucluster19,luhman_tau18}. Additionally, the presence of a distributed, older disk-free membership has been identified through spectral youth indicators with ages potentially as old as $\sim$20\,Myr \citep{taueco1}, and confirmed with variability measurements with time-series photometry (e.g., \citealt{trevorV1298_1}). It is thus possible that comparing two random Taurus stars may result in an age mismatch. 
This is unlikely to be the case for HP~Tau/G2, HP~Tau/G3 and FF~Tau, which are likely coeval, potentially bound, and not associated with a deep column of gas or dust. Additionally, \citet{galli_taucluster19} place these systems in a single Taurus sub-cluster. The discrepancies in age seen in this coeval test-case are largely mirrored for the wider Taurus population. \citet{Kraus_hill09} find that HR-diagram positions of single Taurus stars show a similar mass-age dependence between G and M-type stars.  

The age difference between the G-type HP~Tau/G2, the primaries of these binary systems, and the cooler secondaries is most likely the same model discrepancy in age as a function of stellar mass observed previously observed in young populations, including Taurus \citep{kraushill09}, the somewhat older 10\,Myr  population in Upper Scorpius \citep{preibisch02,pecaut12,dmys1},  and more distance clusters such as NGC~2264 \citep{park00} and the Orion Nebula Cluster \citep{hillenbrand97}, and is attributable to either a luminosity underestimation or temperature overestimation at a particular mass and pre-main sequence age in the model tracks.  An underestimation of model luminosities at a given mass and age of 0.1-0.2\,dex, or corresponding overestimation in model effective temperature of 100-300\,K would account for the age difference between the three Taurus binary systems and HP~Tau/G2. This is consistent with the discrepancies observed in the 10\,Myr old Upper Scorpius population \citep{pecaut12,kraususcoctio5,dmys1}, and also the older pre-main sequence field binary LSPM1314 \citep{dupuy16_lspm1314}.

\section{Implications for the Age and Star Formation History of Taurus}
\label{sectimplications}
There is evidence for a distributed population of slightly older (10-20\,Myr) stars surrounding the Taurus clouds (e.g., \citealt{taueco1}) that formed in a previous epoch of star formation, much like the Sco-Cen-Ophiuchus complex in the south. The currently highly incomplete sample of this population suggests a very low disk fraction, implying that most of these objects have undergone disk dissipation. FF~Tau, HP~Tau/G3, Hubble~4, and HP~Tau/G2 do not show evidence of a gaseous circumstellar disk in the near-IR or at 10-30\,$\mu$m \citep{andrews05,luhman06,furlan06}, though this is not particularly indicative of age. Binary systems undergo disk dissipation on a much shorter timescale than single stars \citep{kraus12_disk}, and so the lack of an observable IR excess for the three binary systems is not inconsistent with their youth. In the case of HP~Tau/G2, because it is a G2-type star, it is not expected to still possess its primordial dust disk \citep{luhman10_taudisk}. At 10\,Myr, only $\sim$13\% of G-type stars retain a debris disk \citep{carpenter09}, and so the lack of a debris disk around HP~Tau/G2  is again not indicative of age.  The proximity of these systems to the molecular/dust clouds ($\sim$1$^\circ$ or $\sim$2-3\,pc  from cloud filament centre), also imply they are likely not part of a distributed older population, but part of the classical Taurus membership.

There is now significant evidence that the current (and previous) generations of pre-main sequence evolutionary models ($<$20\,Myr) under-predict the ages of convective M-type stars in associations of known age in comparison with higher-mass or earlier-type members \citep{pecaut12,kraususcoctio5,dmys1,jeffries16}. We have demonstrated above that the discrepancy extends to a bound and coeval Taurus multiple system (HP~Tau/G2, HP~Tau/G3AB, and FF~TauAB). The classical age for Taurus is 0-2\,Myr \citep{luhman09,kraushill09}, and is based on the HR-diagram positions of the K/M-type population that make up most of the Taurus census. This potential systematic offset in low-mass model stellar ages suggests that Taurus may be older than the classical age, by up to a factor of two, though the introduction of additional physics like magnetic fields might be starting to resolve this discrepancy (e.g., \citealt{feiden16_usco}). This may also be the case for other star-forming regions age dated solely on the basis of HR-diagram positions of K/M-types stars using the current evolutionary models, and could further propagate to the inferred durations of earlier stages of protostellar collapse \citep{evans09}. 

%{\bf Because Taurus has been age-dated using only lower-mass M-type stars, this suggests that Taurus may be somewhat older than the canonically quoted age.} This may also be the case for other star-forming regions age dated solely on the basis of HR-diagram positions of K/M-types stars using the current evolutionary models. We expect that the ages for these groups will require revision to older ages in the future.

%Taurus and its associated clouds are a region of ongoing star formation \citep{taurus_handbook}. Indeed, the presence of Class 0/1 protostars (e.g., \citealt{kenyon95,luhman10})  in the thickest region of the molecular clouds are strong evidence for an ongoing, distributed star-formation event. The classical age for Taurus is 0-2\,Myr \citep{luhman09,kraushill09}, and is based on the HR-diagram positions of the majority M-type population that make up most of the Taurus census. Given that M-type stars on the pre-main sequence appear to have underestimated ages (due to model discrepancies) of up to a factor of 2, the Taurus population age may be closer to $\sim$5\,Myr. This implies that star-formation has been ongoing for at least 5\,Myr in the Taurus region, and potentially longer if a distributed population of older stars around Taurus can be clearly identified.
%These results have further implications for all young association and star-forming regions age dated on the basis of 

\section{Summary}
\label{sectSummary}

We have presented precise astrometric orbits and HST WFC3 photometry of three early K/M-type binary systems in the Taurus-Auriga star-forming region. Using the existing radio parallaxes for these systems, we determine system dynamical masses of $\sim$1-5\% for all three systems, and fit multi-band photometry and spectra to model atmospheres to determine component temperatures and luminosities. We then compared these observations to model evolutionary tracks to determine estimates of component masses and ages.  In summary, we conclude that:

\begin{itemize}

\item The model isochronal ages derived from comparison to evolutionary models \citep{bhac15,chen14_padova,dotter08} for the three binary systems give ages in the range $\sim$1-3\,Myr, which differs significantly from the age of the G2-type star HP~Tau/G2, which is physically associated with HP~Tau/G3 and FF~Tau and thus provides a coeval test for the models.

\item The component model ages from temperatures and luminosities for the lower-mass companions to HP~Tau/G3 and FF~Tau are systematically younger than the corresponding primary components, suggesting a potential continuation of this trend to lower masses.

\item The model age discrepancy corresponds to the model luminosities being under-predicted by 0.1-0.2\,dex, or the models temperatures being too hot by 100-300\,K at a given pre-main sequence age and mass. This discrepancy is consistent with previous binary star results and pre-main sequence HR-diagram age estimation trends with stellar mass for several young populations.

\end{itemize}

\section*{Acknowledgments}
A.C.R was supported as a 51 Pegasi b Fellow though the Heising-Simons Foundation. T.J.D. acknowledges research support from Gemini Observatory. Some of the data presented herein were obtained at the W. M. Keck Observatory, which is operated as a scientific partnership among the California Institute of Technology, the University of California and the National Aeronautics and Space Administration. The Observatory was made possible by the generous financial support of the W. M. Keck Foundation. The authors wish to recognize and acknowledge the very significant cultural role and reverence that the summit of Maunakea has always had within the indigenous Hawaiian community. We are most fortunate to have the opportunity to conduct observations from this mountain. This work was based on observations made with the NASA/ESA Hubble Space Telescope, obtained from the data archive at the Space Telescope Science Institute. STScI is operated by the Association of Universities for Research in Astronomy, Inc. under NASA contract NAS 5-26555. We also thank the anonymous referee for providing insightful and constructive comments.

\bibliographystyle{apj}
\bibliography{master_reference}

\end{document}